\documentclass[a4paper,fleqn,usenatbib]{mnras}

\usepackage{newtxtext,newtxmath}
\usepackage[T1]{fontenc}
\usepackage{ae,aecompl}

\usepackage{graphicx}	% Including figure files
\usepackage{amssymb}	% Extra maths symbols
\title[MCMC population synthesis of pulsars]{Markov chain Monte Carlo population synthesis of single radio pulsars in the Galaxy}

\author[Marek Cie\'{s}lar et al.]{
Marek Cie\'{s}lar,$^{1}$\thanks{E-mail: mcie@camk.edu.pl}
Tomasz Bulik,$^{2}$
Stefan Os{\l}owski$^{3,4,5}$
\\
$^{1}$Nicolaus Copernicus Astronomical Center, Polish Academy of Sciences, Bartycka 18, 00-716 Warsaw, Poland\\
$^{2}$Astronomical Observatory, University of Warsaw, Al Ujazdowskie 4, 00-478 Warsaw, Poland\\
$^{3}$Centre for Astrophysics and Supercomputing, Swinburne University of Technology, Hawthorn, Victoria 3122, Australia\\
$^{4}$Fakult\"{a}t f\"{u}r Physik, Universit\"{a}t Bielefeld, Postfach 100131, 33501 Bielefeld, Germany\\
$^{5}$Max-Planck-Intitut f\"{u}r Radioastronomie, Auf dem H\"{u}gel 69, 53121 Bonn, Germany\\
}

\date{Accepted XXX. Received YYY; in original form ZZZ}

\pubyear{2020}
     
\begin{document}
\label{firstpage}
\pagerange{\pageref{firstpage}--\pageref{lastpage}}
\maketitle

% The almost non-sarcastic abstract of the paper
\begin{abstract}
%The abstract should briefly describe the aims, methods, and main results of the paper.
%It should be a single paragraph not more than 250 words (200 words for Letters).
%No references should appear in the abstract.
    We present a model of evolution of solitary neutron stars, including
    spin parameters, magnetic field decay, motion in the Galactic
    potential and birth inside spiral arms. We use two parametrizations of
    the radio-luminosity law and model the radio selection effects.
    Dispersion measure is estimated from the recent model of free electron
    distribution in the Galaxy (YMW16). Model parameters are optimized
    using the Markov Chain Monte Carlo technique. The preferred model has
    a short decay scale of the magnetic field of $4.27^{+0.4}_{-0.38}$ Myr. However, it has
    non-negligible correlation with parameters describing the pulsar radio luminosity.
    Based on the best-fit model, we predict that the Square Kilometre
    Array surveys will increase the population of known single radio
    pulsars by between 23 and 137 per cent. The Indri code used for
    simulations is publicly available to facilitate future population
    synthesis efforts.

\end{abstract}

%%%%%%%%%%%%%%%%% Introduction %%%%%%%%%%%%%%%%%%%
\section{Introduction}
Evolution of neutron stars (NS) has been a subject of intense studies in the past. 
These objects are primarily observed as radio pulsars but can also be seen in other 
bands like the X-rays, gamma rays as well as in the optical range.
%(\citet{1988Natur.334..684V}, \citet{2017AstL...43..175S}, and \citet{2017arXiv170208280H}). 
There have been numerous efforts to model the radio population. Most notably the works 
of \citet{Narayan1990}, \cite{2006ApJ...643..332F}, \cite{2007Ap&SS.309..245G} 
then \citet{2008MNRAS.388..393K}, \citet{2009MNRAS.395.2326K}, \citet{2011MNRAS.413..461O} 
and in recent years \citet{2013MNRAS.434.1387L}, \citet{2014MNRAS.443.1891G}, and \citet{2014MNRAS.439.2893B}. 
For an in-depth review of population synthesis efforts see \citet{2007PhyU...50.1123P} 
and \citet{2011heep.conf...21L}.

We base our motivation to revisit the radio population of pulsars on the improved
model of the electron density in the Galaxy \citet{2017ApJ...835...29Y}, and the 
availability of Markov Chain Monte Carlo (MCMC) to explore the multidimensional 
parameter space due to the extended computational power.
We restrict our analysis to the evolution of the single pulsars from their birth 
in a supernova explosion to the moment they no longer can be detected in the radio 
waveband. We do not consider binary evolution and interactions therefore treatment 
of millisecond pulsars is beyond the scope of this paper. We do not simulate the 
full stellar and binary evolution that leads to formation of pulsars, such as was 
done by  \citet{2008MNRAS.388..393K}, \citet{2009MNRAS.395.2326K} and 
\citet{2011MNRAS.413..461O} and therefore we start with pulsars progenitor 
distribution as an input parameter to the simulation.
We provide the {\em Indri} source code\footnote{The code can be obtained from the 
{\em GitHub} repository \url{http://github.com/cieslar/Indri}} with an intent 
that one can fully reproduce our results upon access to a small cluster\footnote{At 
the time of writing we define such machine as an approximately $200$-cores cluster.}, 
expand the scope of the simulation, use different data cuts or jump-start further development. 

The paper is arranged as follows: in section \ref{sec:model} we explain the Galactic model, the kinematics of pulsar 
population, the evolution of the pulsars period and the magnetic field, luminosity 
models, the selection effects as well as the mathematical representation of the model, 
in section \ref{sec:obs} we describe the construction of the likelihood of the 
model upon comparison with survey data and describe the implementation of the 
Mertropolis-Hasting {\em MCMC} method, in section \ref{sec:results} we present 
the results obtained in the simulation, we discuss them in the section \ref{sec:discussion}
, and we summarize in section \ref{sec:conclusions}.

%%%%%%%%%%%%%%%%% Model %%%%%%%%%%%%%%%%%%%%%%%%%%
\section{The Model}\label{sec:model}
There are two broadly independent parts that are needed to describe the 
evolution of NSs. The first part is connected with the dynamical evolution of NSs 
in the gravitational potential of the Milky Way, and the second describes the 
intrinsic evolution in time of each neutron star as a radio pulsar. The model is 
roughly following the one presented by \citet{2006ApJ...643..332F}. In the following section 
we concentrate on the differences between our model and \citet{2006ApJ...643..332F}, while 
the identical components are presented in the Appendix \ref{sec:FK06reimplementation}. 
We assume that the pulsars birth time has a uniform distribution. We model the evolution over a 
period of $t_{\rm max}=50\,{\rm Myr}$. It is important to note that the 
characteristic age $\tau=P/2\dot{P}$ can reach much higher values than 
$t_{\rm max}$ because of the magnetic field decay (see discussion in 
\ref{sec:taudisc}).

%%%%%%%%%%%%%%%%% Galaxy %%%%%%%%%%%%%%%%%%%%%%%%%
\subsection{The Milky Way}
\label{sec:Galaxy}

%\footnote{\url{http://mathworld.wolfram.com/SpherePointPicking.html}}. 
\subsubsection{The equation of motion - integration method}
We use the {\em Verlet} method \citep{1967PhRv..159...98V} to propagate pulsars 
through the Galactic potential. Following a  Monte Carlo experiment (simulated 
motion of few millions of random pulsars), we found that the maximal time step 
can not exceed $dt_{\rm max}=0.1\,{\rm Myr}$ in order to limit the loss of 
the total energy to $1\%$ due to the numerical errors. The actual time step 
$dt_{\rm act}$ is lower then $dt_{\rm max}$ and it's equal to:
\begin{equation}
dt_{\rm act}=\frac{t_{\rm age}}{\frac{t_{\rm age}}{dt_{\rm max}}+1}
\end{equation}
We discard pulsars which are more than $35\,\rm{kpc}$ away from the Galactic centre at the end of the simulation.

%%%%%%%%%%%%%%%%%%%%%%%%%%%%%%%%%%%%%%%%%%%%%%%%%%
%%%%%%%%%%%%%%%%% NS Physics %%%%%%%%%%%%%%%%%%%%%
\subsection{The neutron star physics}\label{sec:Phys}
We assume constant values for the radius ($R_{\rm NS}=10\,{\rm km}$), 
the mass ($M_{\rm NS}=1.4\,{\rm M}_\odot$) and the moment of inertia 
($I_{\rm NS}=10^{45}\,{\rm g\,cm}^2$) of each neutron star.

\subsubsection{The magnetic field decay}
Following \citet{2011MNRAS.413..461O} we assume that the magnetic field decays due
to the Ohmic dissipation. For recent advanced we refer to the work of 
\citet{2015AN....336..831I}, though we simplify the time dependence of the decay to
an exponential function. The decay model is parametrised by the time-scale $\Delta$: 
\begin{equation}
B(t)=(B_{\rm init}-B_{\rm min})\exp{\left( \frac{-t}{\Delta} \right) } + B_{{\rm min}}
\label{eq:Bdecay}
\end{equation} 
To be consistent with our previous work \citet{2011MNRAS.413..461O}, and with 
\citet{2008MNRAS.388..393K}, we use the minimum value of the magnetic field
given by \citet{ 2006MNRAS.366..137Z}. We draw it from a log-uniform distribution:
\begin{equation}
10^{7}\,{\rm G} < B_{{\rm min}} < 10^{8}\,{\rm G}
\end{equation}
The results do not depend on the choice of $B_{\rm{min}}$ since the pulsars with such 
small magnetic field are not included in the comparison sample as they no longer emit in radio.

\subsubsection{The evolution in time}
The boundary conditions for the pulsars evolution are their initial and final 
magnetic field strength $B_{\rm init}$, $B_{\rm min}$ as well the spin period 
at birth $P_{\rm init}$. To obtain a set of values $P$ and $B$ at the time 
$t_{\rm age}$ we integrate the radiating magnetic dipole (equation 
\ref{eq:BPPdotRealtion}) by supplying it with the magnetic field decay 
(equation \ref{eq:Bdecay}):
\begin{equation}
    P(t_{{\rm age}}) = \frac{1}{\eta} \sqrt{ 2\int_{0}^{t_{{\rm age}}}\left(B^2(t){\rm d}t\right)  + P_{\rm init}^2}
\end{equation}
where $\eta\simeq 3.2 \times 10^{19}\,{\rm G}\,{\rm s}^{-1/2}$.
\begin{equation}
\begin{split}
    P(t_{{\rm age}}) = \left(\frac{1}{\eta^2}\left( \left(B_{\rm init}-B_{{\rm min}}\right)^2\left(1-{\rm e}^{-\frac{2t_{\rm age}}{\Delta}}\right){\Delta}\right.\right.+     \\
    \left.\left. + 4B_{{\rm min}}\left(B_{\rm init}-B_{{\rm min}}\right)\left(1-{\rm e}^{-\frac{t_{\rm age}}{\Delta}}\right){\Delta} + 2B_{{\rm min}}^{2}t_{\rm age}\right) + P_{\rm init}^{2}\right)^{\frac{1}{2}}
\end{split}
\end{equation}
We obtain $\dot{P}$ by inserting $P(t_{{\rm age}})$ in equation \ref{eq:BPPdotRealtion}.

%%%%%%%%%%%%%%%%%%%%%%%%%%%%%%%%%%%%%%%%%%%%%%%%%%
\subsection{Radio Properties}\label{sec:RadioBias}
\subsubsection{The phenomenological radio luminosities}
\label{seq:RadioLuminosities}
Since the first pulsar detection \citep{1969Natur.224..472H}, their radio 
emission process is still in debate \citep{2015SSRv..191..207B}. In our work we
assume a simple model of pair creation. Though, due to the phenomenological treatment 
of the luminosity it does not add any constraints, it is of significance only 
while considering the {\em death lines} (see section \ref{sec:deatharea}). In this
paper we use two different {\em a priori} assumptions about the radio luminosity.
\paragraph*{The two-parameter power law.} 
The general approach to describe the phenomenological relation between the $P$ 
period, period derivative $\dot{P}$ and the radio luminosity $L_\nu$ at frequency
$\nu$ is a power law with two parameters $\alpha,\beta$ and a scaling factor
$\gamma$ see e.g. \citet{2006ApJ...643..332F} and \citet{2014MNRAS.439.2893B}:
\begin{equation}
    L_{400,\rm{p-l}}=\gamma P^{\alpha}\dot{P}_{15}^{\beta}\,{\rm mJy}\times{\rm kpc}^2,
    \label{eq:Lpowerlaw}
\end{equation}
for the observational frequency $\nu=400\,{\rm MHz}$.\\ 

\paragraph*{The rotational energy power law.} A more restricted model is the
one in which the luminosity is proportional to the rotational energy loss 
see e.g. \citet{Narayan1990}:
\begin{equation}
    L_{\rm{rot}}\equiv -\dot{E}_{\rm {rot}}=\frac{4\pi^2\dot{P}}{P^3},
\label{eq:Lrot}
\end{equation}
\begin{equation}
    L_{400,\rm{rot}}=\gamma\left(\dot{P}_{15}^{\frac{1}{3}}P^{-1}\right)^\kappa \,\rm{mJy}\times\rm{kpc}^2.
\end{equation}
We include the correction 
$L_{\rm corr}$ proposed by \citet{2006ApJ...643..332F} and adopted by 
    \citet{2014MNRAS.439.2893B} to both radio flux laws (eq. \ref{eq:Lpowerlaw} and \ref{eq:Lrot}):
\begin{equation}
    \log L_{400}=\log\left( L_{400,\rm{rot/p-l}}\right)+L_{\rm{corr}}
\end{equation}
The $L_{\rm{corr}}$ is randomly drawn from the normal distribution with $\sigma_{corr}=0.8$ 
and accounts for spread of observed population around any parametric models of radio luminosity.
We assume that the radio spectrum can be described by a power law:
\begin{equation}
L_{\nu}=L_{\nu_{0}}\left( \frac{\nu}{\nu_{0}} \right)^{\alpha_{{\rm sp}}}
\end{equation}
with the spectral index $\alpha_{{\rm sp}}=-1.4$ \citep{2000A&AS..147..195M}.  
Pulsar emission is highly anisotropic. In order to  model the geometry of the beam 
from a pulsar we incorporate the beaming factor following \citet{1998MNRAS.298..625T}.
For a pulsar with the period $P$ we calculate the beaming fraction $f(P)$ in percent: 
\begin{equation}
f(P) = 9\times \left(\log{\frac{P}{10 \mathrm s }}\right)^{2}+3
\end{equation}
and determine the visibility of each pulsar assuming random orientation.
%%%%%%%%%%%%%%%%%%%%%%%%%%%%%%%%%%%%%%%%%%%%%%%%%%

\subsubsection{Death lines -- death areas}
\label{sec:deatharea}
\begin{figure}
	\includegraphics[width=\columnwidth]{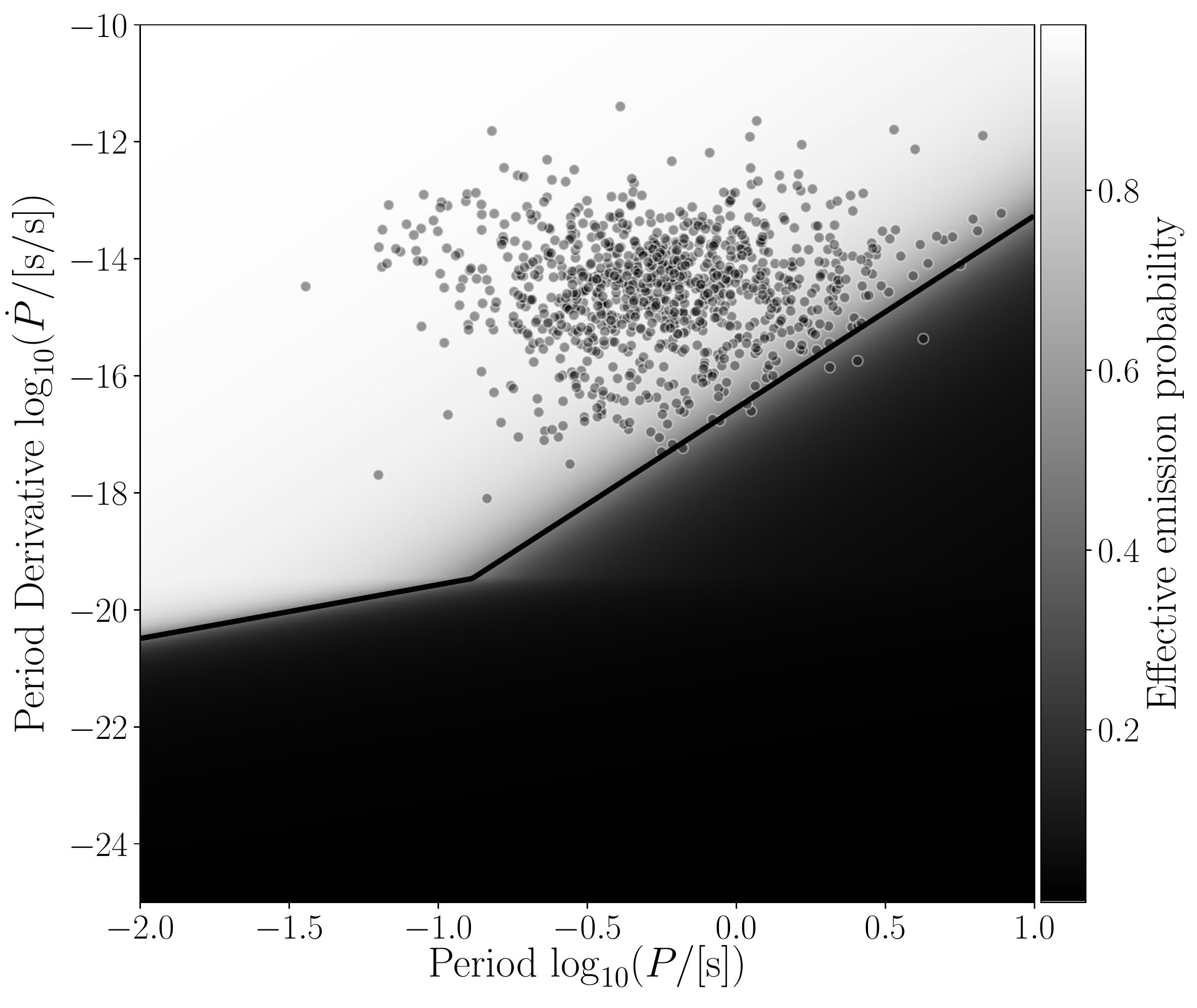}
	\caption{Death Area ($\mathfrak{D}_{{\rm Area}}$) - the effective emission 
    probability. Solid dark lines - canonical death lines ($\mathfrak{D}_{{\rm Line}}$) 
    by \citet{1994MNRAS.267..513R}. Grey points - sub selection of ATNF catalogue used in the simulation.}
	\label{fig:DeathArea}
\end{figure}
In the canonical emission process \citep[see][]{1967Natur.216..567P,1968Natur.218..731G} the radio waves are emitted due to the $e^\pm$ pair 
creation and their acceleration and cascade creation in the presence of strong 
magnetic field. The pulsars radio emission process stops when the processes 
cannot be sustained \citep{1994MNRAS.267..513R}.
These so-called {\em death-lines} are defined as:
\begin{equation}
\log{\dot{P}} = 3.29 \times \log{P} -16.55,
\end{equation}
\begin{equation}
\log{\dot{P}} = 0.92 \times \log{P} -18.65.
\end{equation}
Any pulsar crossing them during its evolution is considered radio inactive.
However, such model contradicts the observations as a number of pulsars lie 
below these lines. This discrepancy can be attributed to the fact that death 
lines are devised for a specific structural model and parameters of the neutron star.
Similarly to \citet{2002ApJ...568..289A}, we propose a phenomenological function to smooth the death lines into a continuous
death area (see Figure \ref{fig:DeathArea}). We propose a following formula: 
\begin{equation}
\mathfrak{D}_{{\rm Area}}(P,\dot{P}) = \frac{1}{\pi} \arctan\left( \frac{\log P - \mathfrak{D}{_{\rm Line}}(\log \dot{P})}{\Psi}\right) + 0.5 .
    \label{eq:Death}
\end{equation}
The value of $\Psi$ parameter is set to $0.2$ in order for the probability of 
radio activity to change in the range of $\rm{d}\log P \sim 1$. 
\subsubsection{The dispersion measure}
We compute the dispersion measure ${\rm DM}$ for each pulsar in the model
population using the new and updated model of the electron density in the
Milky Way \citep[][YMW16]{2017ApJ...835...29Y}. The {\em Indri} code can also use the 
NE2001 model \citep[see][]{2002astro.ph..7156C,2003astro.ph..1598C}.

%%%%%%%%%%%%%%%%%%%%%%%%%%%%%%%%%%%%%%%%%%%%%%%%%%
\subsection{The computations}
\subsubsection{The mathematical representation}
\label{PSRDensPropab}
To mathematically represent the model we construct pulsar density in a three-dimensional space and smooth it with a Gaussian kernel.
This {\em comparison space} is spanned by 
the period $P$, the period derivative $\dot{P}$ and the flux at 
$1400\,{\rm MHz}$, $S_{1400}$ (shortened to $S$ hereafter).
The Gaussian averaged number of pulsars at a particular point (specified by
indices $k,l,m$) of the {\em comparison space} ${\log}P_k$-${\log}\dot{P}_l$-${\log}S_m$ 
is expressed by:
\begin{equation}
    \begin{split}
        \bar{\rho}_{klm} = \sum\limits_{b}^{{\rm PSR}} \frac{1}{\left(2\pi\sigma_{{\rm cs}}\right)^{\frac{3}{2}}}
        \exp\left(-\frac{\left({\log}P_{b} - {\log}P_{k}\right)^{2}}{2\sigma_{{\rm cs}}^{2}}\right)\cdot\\
        \exp\left(-\frac{\left({\log}\dot{P}_{b} - {\log}\dot{P}_{k}\right)^{2}}{2\sigma_{{\rm cs}}^{2}}\right)
        \exp\left(-\frac{\left({\log}S_{b} - {\log}S_{k}\right)^{2}}{2\sigma_{{\rm cs}}^{2}}\right),
    \end{split}
\end{equation}
where $\sigma_{{\rm cs}}$ is equal to $0.2$. The particular value of the meta-parameter $\sigma_{{\rm cs}}$ 
has been heuristically chosen based on the behaviour of the model. Too low and the algorithm (described in section \ref{seq:MCMC})
would never converge. Too large and the model would reflect and find only the maximum of the three-dimensional distribution in the $\log P$-$\log \dot{P}$-$\log S$ space.
To normalise the $\bar{\rho}_{klm}$
we use the sum $R$ over all relevant points (located near observations):
\begin{equation}
    R=\sum\limits_{l,k,m}\bar{\rho}_{klm}
\end{equation}
And then, construct the normalised, Gaussian averaged, pulsar density:
\begin{equation}
    \rho_{klm} = \frac{1}{R}\bar{\rho}_{klm}
\end{equation}
For the ease of notation we re-index the $k,l,m$ indices with single $i$-index 
traversing all combinations of the $k,l,m$ set. So that $\rho_{i}{\coloneqq}\rho_{klm}$ 
represents a distinct point in the ${\log}P_k$-${\log}\dot{P}_l$-${\log}S_m$ space.

\subsubsection{The performance of the evolutionary code}\label{sec:performance} 
We have found that the main performance bottleneck in our computations is the 
evaluation of the dispersion measure in the YMW16 model. The code  provided by 
\citet{2017ApJ...835...29Y}\footnote{We use the version $1.2.2$ from 
\url{www.xao.ac.cn/ymw16/}} was not intended to be a part of a high performance 
computation and thus, we faced a choice. We could scale back the computation and 
abandon the {\em Monte Carlo} approach of the parameter search. Or we might make 
the galactic part of the model static losing the ability to test supernova kicks 
and initial position assumptions. We chose the latter option. 
The resulting algorithm is executed in two steps:
\begin{itemize} 
\item[(i)] We simulate the motion in the galactic potential (as described in the 
    \ref{sec:Galaxy} section). The goal is to have one million neutron stars that 
        are in the sky-window of the Parkes Survey. This number of pulsars is 
        chosen for practical, computational reasons. We call this set of stars 
        {\em the geometrical reference population}. 
    \item[(ii)] We take the {\em geometrical reference population} (the age, 
        dispersion measure and distance) and use it as an input for physical 
        computation (the \ref{sec:Phys} section). We use each NS from the 
        {\em geometrical reference population} $5$ times, i.e. we place five 
        different model pulsars at each location, so that they have the same 
        position in the sky and the same dispersion measure. The evolution 
        computations finish with the radio-visibility test (the \ref{sec:RadioBias} 
        section). We check whether the pulsar is beaming towards Earth and if it is 
        emitting radio waves according to the death area criteria. If both conditions 
        are satisfied we compute the luminosity $L_{400}$ and the detected flux 
        on Earth. To finish the test we check if the pulsars flux is higher then 
        his minimal detectable flux. The population that satisfies the 
        radio-visibility test is called {\em the model population}. This step 
        ends with the computations of {\em the likelihood} statistic in the 
        comparison space (see the \ref{sec:Likelihood} section).
\end{itemize}
The first step is done only once while the second step is used for the intensive 
{\em Monte Carlo} computations described in the following section. Such scheme 
allows us to investigate the model by using a population of five million pulsars.
We note that should the YMW16 model be rewritten in computationally efficient way, 
it would be possible to carry out the simulation with the inclusion of a 
parametrisation of the initial positions, the SN kicks, and the Galactic potential. 

%%%%%%%%%%%%%%%%%%%%%%%%%%%%%%%%%%%%%%%%%%%%%%%%%%
%%%%%%%%%%%%%%%%%%%%%%%%%%%%%%%%%%%%%%%%%%%%%%%%%%
%%%%%%%%%%%%%%%%%%%%%%%%%%%%%%%%%%%%%%%%%%%%%%%%%%
\section{Comparison with Observations}\label{sec:obs}

%%%%%%%%%%%%%%%%%%%%%%%%%%%%%%%%%%%%%%%%%%%%%%%%%%
\subsection{The observations}
\label{sect:ATNFCatalogue}
For the verification, we compare our model with a subset of the Australia Telescope National
Facility Pulsar Catalogue\footnote{version $1.54$, \url{http://www.atnf.csiro.au/research/pulsar/psrcat}}
\citep{2005AJ....129.1993M}.
We perform the following cuts to select an unbiased sample of pulsars:
\begin{itemize}
\item [(i)] we preselect single pulsars with measured necessary parameters 
    ($P$, $\dot{P}$, $S_{1400}$, $l$, $b$, and ${\rm DM}$), 
\item [(ii)] we choose only the pulsars that have been observed by the Parkes 
    Multibeam Survey \citep{2001MNRAS.328...17M},
\item [(iii)] since we focus on the evolution of single pulsars we neglect the 
    potentially recycled ones by requiring the inferred surface magnetic field to
    be greater then $10^{10}\,{\rm G}$.
\end{itemize}
With these cuts we obtained a subset of $969$ pulsars. In order to be consistent,
we perform the second and third cuts as  throughout {\em the model population} as well.

%%%%%%%%%%%%%%%%%%%%%%%%%%%%%%%%%%%%%%%%%%%%%%%%%%
\subsubsection{The comparison between model and observations}\label{sec:ModelObsProb}
The pulsars density described in section \ref{PSRDensPropab} can be expressed for both
the model ($\rho \to m$) or the observations ($\rho \to o$). 
For a given $i$-th point of the comparison space, we compare the model $m_{i}$ pulsar
density with the observed $o_i$ pulsar density. Using the central limit theorem, we 
assume that the probability that the measured density $o_i$ has its value given the 
model density $m_i$ is described by a normal distribution:
\begin{equation}
\mathcal{P}_i(\bar{\theta})=\mathcal{P}(m_{i}(\bar{\theta}),o_{i}) = \frac{1}{\sqrt{2\pi}}\exp{\left(-\frac{(m_{i}(\bar{\theta})-o_{i})^{2}}{2}\right)}
\label{eq:ModelObsProb}
\end{equation} 
where we denoted the model parameters as $\bar{\theta}$.
For numerical reasons, it more convenient to work with the logarithm of the probability $\mathcal{P}_i$:
\begin{equation}
\ln\mathcal{P}_i(\bar{\theta})=-\ln(\sqrt{2\pi}) -\frac{(m_{i}(\bar{\theta})-o_{i})^{2}}{2}
\end{equation}

\subsubsection{Likelihood}\label{sec:Likelihood}
In order to find optimal parameters for the model we use the likelihood statistic.
In general,
the likelihood $\mathcal{L}$ of $n$ independent variables $x_1,\ldots,x_n$ drawn from
an unknown probability distribution parametrised by $\bar{\theta}$ is expressed by a
joint probability function $f$:
\begin{equation}
\mathcal{L}(\bar{\theta}\,;\,x_1,\ldots,x_n) = f(x_1,x_2,\ldots,x_n\mid\bar{\theta}) 
\end{equation}
The joint probability function $f$ is a product of probability functions $g$:
\begin{equation}
f(x_1,x_2,\ldots,x_n\mid\bar{\theta}) = \prod_{i=1}^n g(x_i\mid\bar{\theta}).
\end{equation}
In our case, due to the finite number of points in the comparison space, the 
independent variables $x_1,\ldots,x_n$ are represented by the points $m_i$ (as 
defined in the section \ref{PSRDensPropab} and \ref{sec:ModelObsProb}). The
probability density function $g$ is represented by $\mathcal{P}$ (equation
\ref{eq:ModelObsProb}):
\begin{equation}
\mathcal{L}(\bar{\theta}) = \prod_{i \in \Omega} \mathcal{P}_i(\bar{\theta})
\end{equation}
where $\Omega$ denotes the set of points at which we calculate the pulsars density $\rho_i$.
For our computation we use the logarithm of the likelihood:
\begin{equation}
\ln\mathcal{L}(\bar{\theta}) = \sum_{i \in \Omega} \ln\mathcal{P}_i(\bar{\theta})
\end{equation}

%%%%%%%%%%%%%%%%%%%%%%%%%%%%%%%%%%%%%%%%%%%%%%%%%%
\subsection{Markov chain Monte Carlo}\label{seq:MCMC}
To find the most probable parameters of the model we use the {\em Markov chain 
Monte Carlo} technique (MCMC). For the discussion of this widely used and 
established concept we refer to the works of \citet{tarantola2005inverse}, 
\citet{mackay2003information} or \citet{2017ARA&A..55..213S}. In our case, we 
use the {\em Metropolis-Hastings} random walk \citep{hastings1970monte} approach 
to construct chains of likelihood values.
At the start of each chain, the parameter vector $\bar{\theta}$ is randomly drawn 
from the whole available parameters subspace (see Table \ref{tab:ParametersConstraints} 
and section \ref{sec:results} for parameters definitions) using a flat distribution 
in each dimension.
During the random walk phase, the new set of parameters is drawn according to the 
normal probability distribution centered at the old set of parameters. The drawing 
is done independently for each $i$-th parameter:
\begin{equation}
P(\theta^{i}_{n},\theta^{i}_{p})=\frac{1}{\sqrt{2 \pi \sigma_{\theta^{i}}^{2}}}\exp \left( \frac{-(\theta^{i}_{n}-\theta^{i}_{p})^2}{2\sigma_{\theta^{i}}^2} \right)
\label{eq:jumpprob}
\end{equation}
where the $\sigma_{\theta^{i}}$ is set to a $\frac{1}{3}$-th of the parameter interval
(for the interval description see Table \ref{tab:ParametersConstraints}). 
To draw parameters whose initial distribution is log-normal, we
replace the value of the parameter with its logarithm in equation \ref{eq:jumpprob}.
If the newly drawn parameter is outside of bounds the procedure is repeated.
Following the methodology presented by \citet{1995JGR...10012431M} we use the 
likelihood-modified step function to decide if the chain will move to the next 
location in the {\em Metropolis-Hastings} algorithm:
\begin{equation}
\mathcal{R}_{pn} = \frac{\mathcal{L}(\bar{\theta}_{n})}{\mathcal{L}(\bar{\theta}_{p})}
\end{equation}
where $p$ and $n$ denote the {\em previous} and {\em next} set of parameters 
$\bar{\theta}$ of a given step. If  $\mathcal{R}_{pn} \geqslant 1$, the jump is 
certain. If it is $\mathcal{R}_{pn} < 1$ then a jump to {\em next} set of parameters 
is done with the probability equal to 
$\mathcal{R}_{pn}$. 
The calculations are repeated until the distribution of chain end-points becomes 
subjectively stable.

\subsubsection{Optimization and verification of the MCMC}
We have learnt that some of the Markov chains converge on the maximum too slowly 
or not at all (should they be initially located too far from the extrema in the
parameter space). This is well known, general problem of MCMC methods. It differs
between algorithms and techniques and can be, depending on the technique, minimized 
to some extent. 
%The usual solution is to employ more sophisticated method (see \citet{gilks1995adaptive}
%and \citet{2013PASP..125..306F}).
Instead of implementing more sophisticated method \citep[see][]{gilks1995adaptive, 2013PASP..125..306F},
we did two kinds of simulations. A general one, with a broad step-size to pinpoint 
the general area of the maximum of the likelihood (as described in previous subsection). And a second one, starting from 
a single point in the vicinity of the maximum of likelihood with four times smaller
step-size (a $\frac{1}{12}$-th of the parameter interval described in Table \ref{tab:ParametersConstraints}). 
. We have run $1000$ chains, each $5000$-links long. We confirmed that the 
chains reach stability by computing the integrated autocorrelation time IAT\footnote{We used the procedure {\em acor} from the \url{https://github.com/dfm/acor} repository.}\citep[see][]{2010CAMCS...5...65G}.
The maximum IAT values, among all marginal parameters distributions, were 279 and 176 links for the {\em power-law} and {\em rotational} model, respectively. 

%%%%%%%%%%%%%%%%%%%%%%%%%%%%%%%%%%%%%%%%%%%%%%%%%%
%%%%%%%%%%%%%%%%%%%%%%%%%%%%%%%%%%%%%%%%%%%%%%%%%%
%%%%%%%%%%%%%%%% Results %%%%%%%%%%%%%%%%%%%%%%%%%
%%%%%%%%%%%%%%%%%%%%%%%%%%%%%%%%%%%%%%%%%%%%%%%%%%
%%%%%%%%%%%%%%%%%%%%%%%%%%%%%%%%%%%%%%%%%%%%%%%%%%
\section{Results}\label{sec:results}
We limited our studies to two models -- the {\em power-law} and the {\em rotational} 
model. They differ in the phenomenological description of the radio luminosity (see 
section \ref{seq:RadioLuminosities}).
To describe them, we use a 8 (for the power-law model) or 7 (for the rotational model) 
parameters listed in Table \ref{tab:ParametersConstraints}.
Four of the parameters are used to describe the initial conditions: distributions of
magnetic fields ($\widehat{B}_{{\rm init}}$, $\sigma_{B_{{\rm init}}}$) and 
periods ($\widehat{P}_{{\rm init}}$, $\sigma_{P_{{\rm init}}}$). One parameter
($\Delta$) is associated with the decay scale of the magnetic field. And the remaining
three ($\gamma$, $\alpha$, $\beta$) in case of the {\em power-law} or two ($\gamma$,
$\kappa$) in the {\em rotational} model describe the radio luminosities.
\begin{table}
\centering
\begin{tabular}{llll}
Parameter & Min value & Max value & Space\\
\hline
    $\alpha$ & $-2.$ & $2.$ & $1$ \\
    $\beta$ & $-0.5$ & $1.$ & $1$ \\
    $\kappa$ & $0.2$ & $1.4$ & $1$ \\
    $\gamma\,{\rm mJy}$ & $10^{-4}$ & $10^{4}$  & $\log$ \\
    $\Delta\,{\rm Myr}$ & $10^{0}$ & $10^{2}$ & $\log$ \\
    $\widehat{B}_{\rm init}\,{\rm G}$ & $10^{12}$ & $10^{13}$ & $\log$\\
    $\sigma_{B_{\rm init}}\,{\rm G}$ & $10^{0.25}$ & $10^{0.7}$ & $\log$\\
    $\widehat{P}_{{\rm init}}\,{\rm s}$ & $0.01$ & $0.6$ & $1$ \\
    $\sigma_{P_{{\rm init}}}\,{\rm s}$ & $0.01$ & $0.6$ & $1$ \\
\hline
\end{tabular}
\caption{The constraints of the parameters. The {\em Space} column indicates whether 
    the parameters axis is linear or logarithmic. It also correspond to the parameters 
    jump probability distribution -- normal or log-normal, respectively. The $\alpha$,
    $\beta$, and $\kappa$ parameters are dimensionless.}
\label{tab:ParametersConstraints}
\end{table}

\begin{figure*}
\centering
\includegraphics[width=\textwidth]{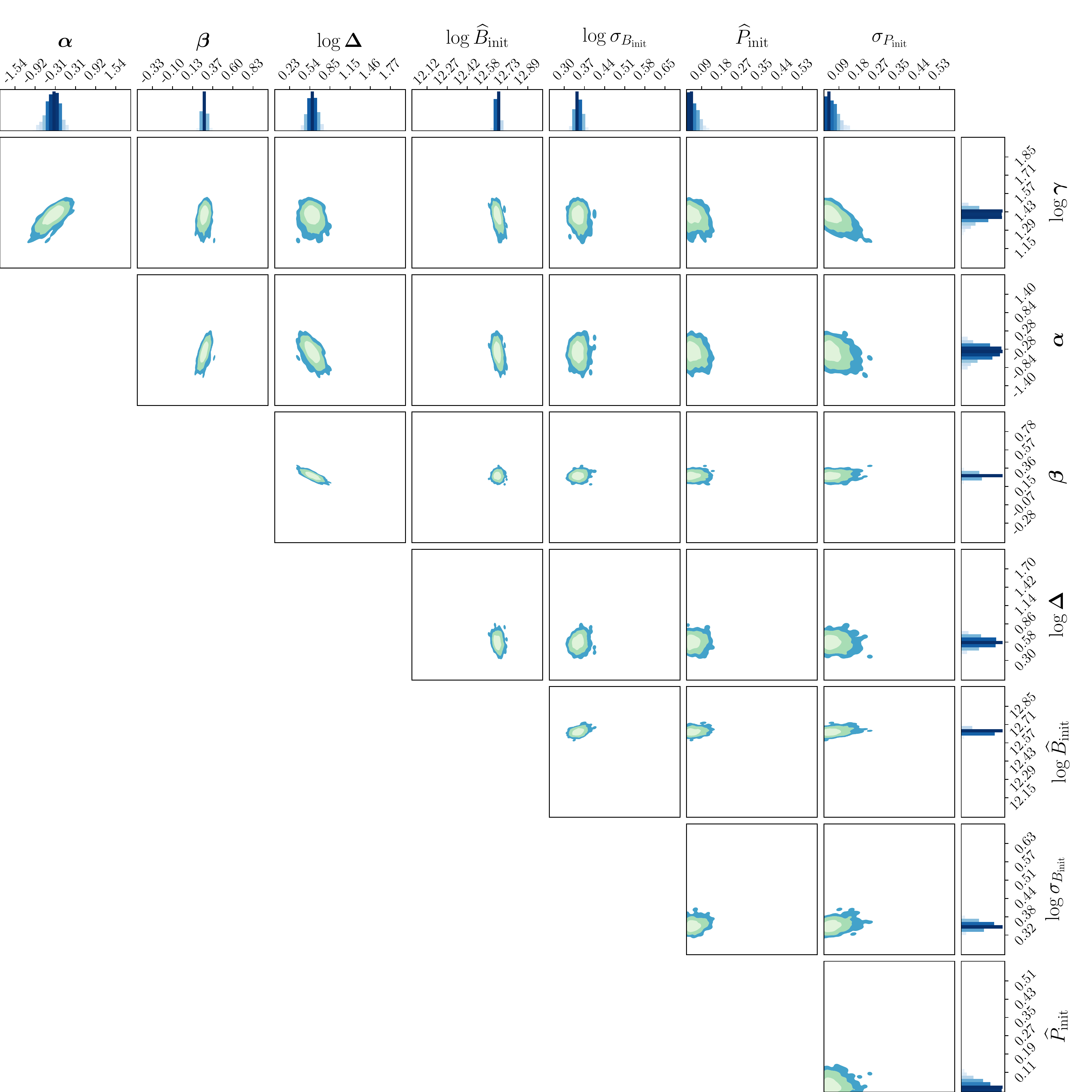}
\caption{The {\em power-law} model -- MCMC marginal parameter space. The 1D marginal 
    distribution express the auto-scaled, normalised probability density ${\rm d}N/(N{\rm d}i)$ 
    where $i$ is appropriate parameter according to the plot. On the 2D contour plot the colours 
    represent the $1,2,3-\sigma$ levels. We constrained the range of the $\gamma$ posterior to zoom in on the populated part of the phase space.} 
\label{fig:ContourPowerlLaw}
\end{figure*}
\begin{figure*}
\centering
\includegraphics[width=\textwidth]{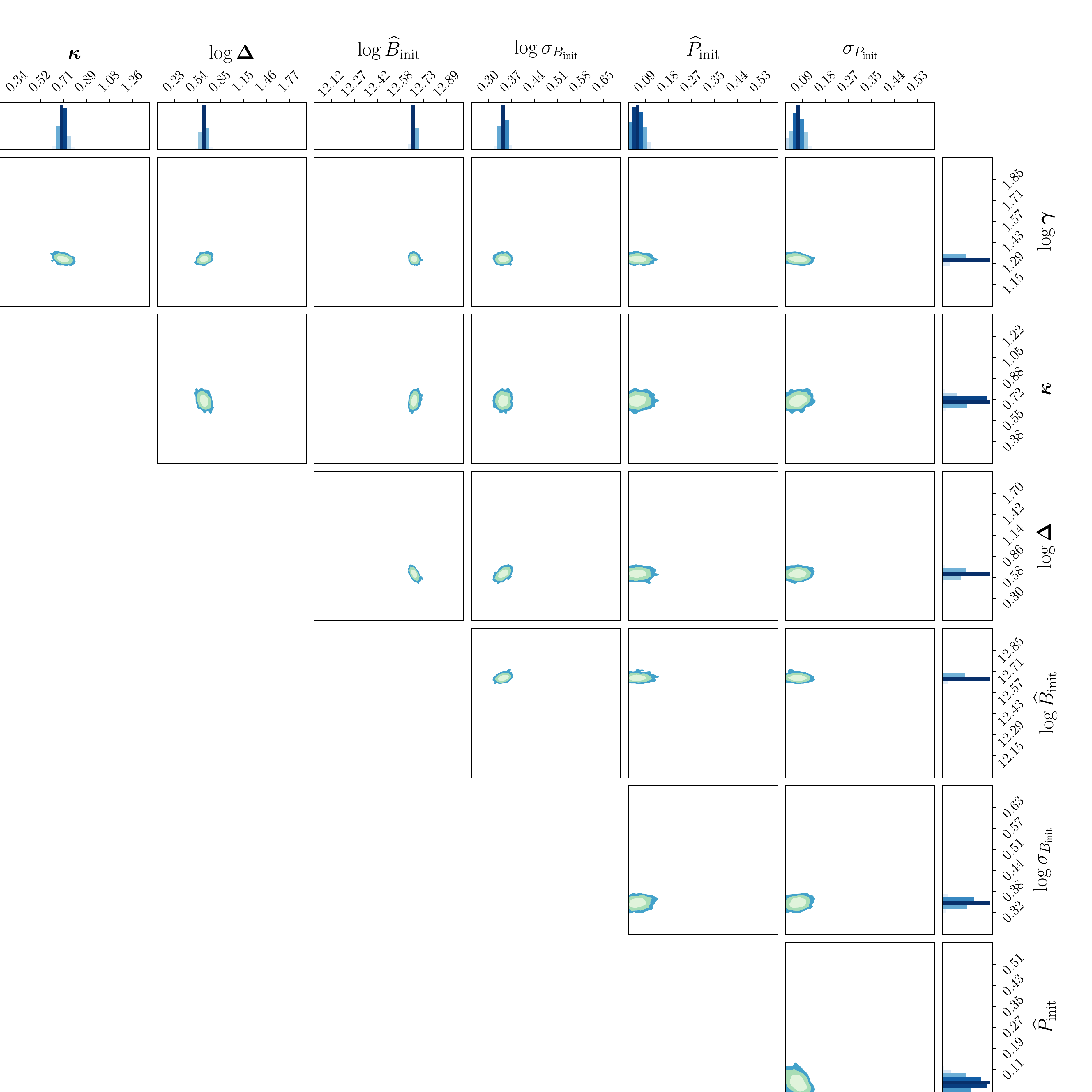}
\caption{The {\em rotational} model -- MCMC marginal parameter space. The 1D marginal 
    distribution express the auto-scaled, normalised probability density ${\rm d}N/(N{\rm d}i)$ 
    where $i$ is appropriate parameter according to the plot. On the 2D contour plot the colours 
    represent the $1,2,3-\sigma$ levels. We constrained the range of the $\gamma$ posterior to zoom in on the populated part of the phase space.} 
\label{fig:ContourRotational}
\end{figure*}
\begin{figure*}
\centering
\begin{tabular}{cc}
The {\em power-law} model & The {\em rotational} model \\
\includegraphics[width=0.48\textwidth]{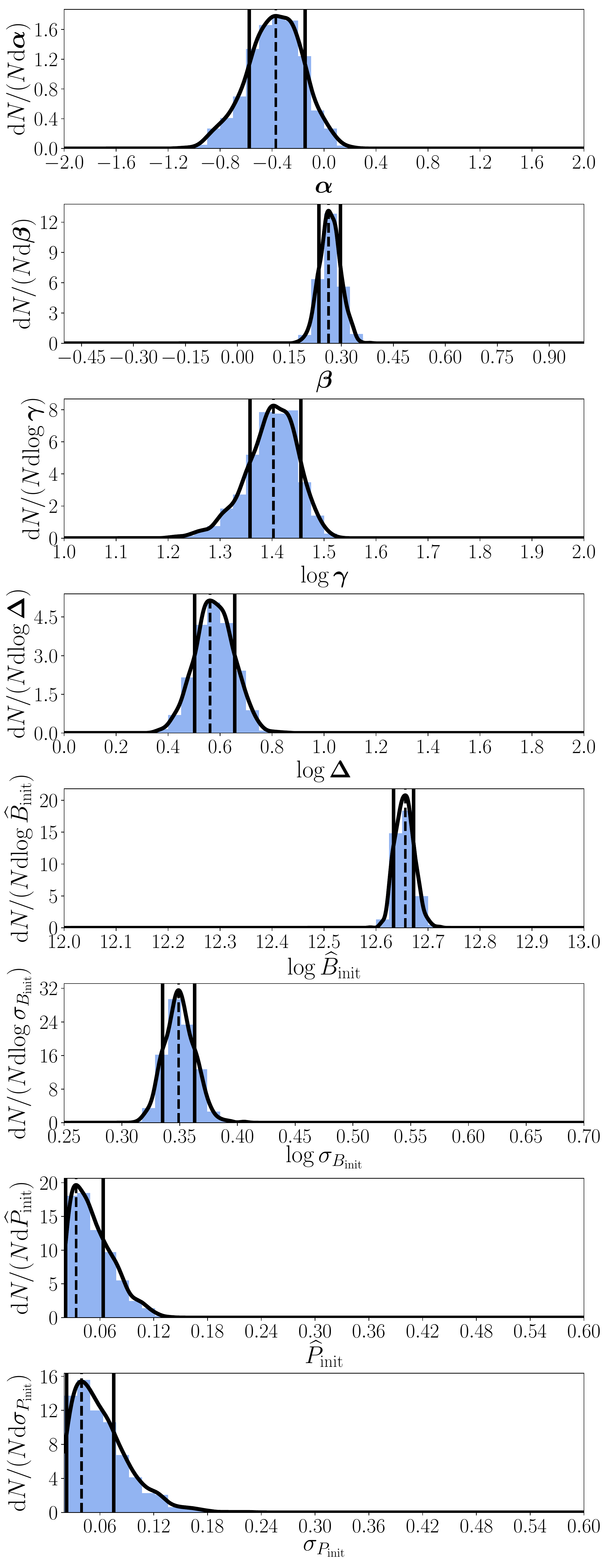} & \includegraphics[width=0.48\textwidth]{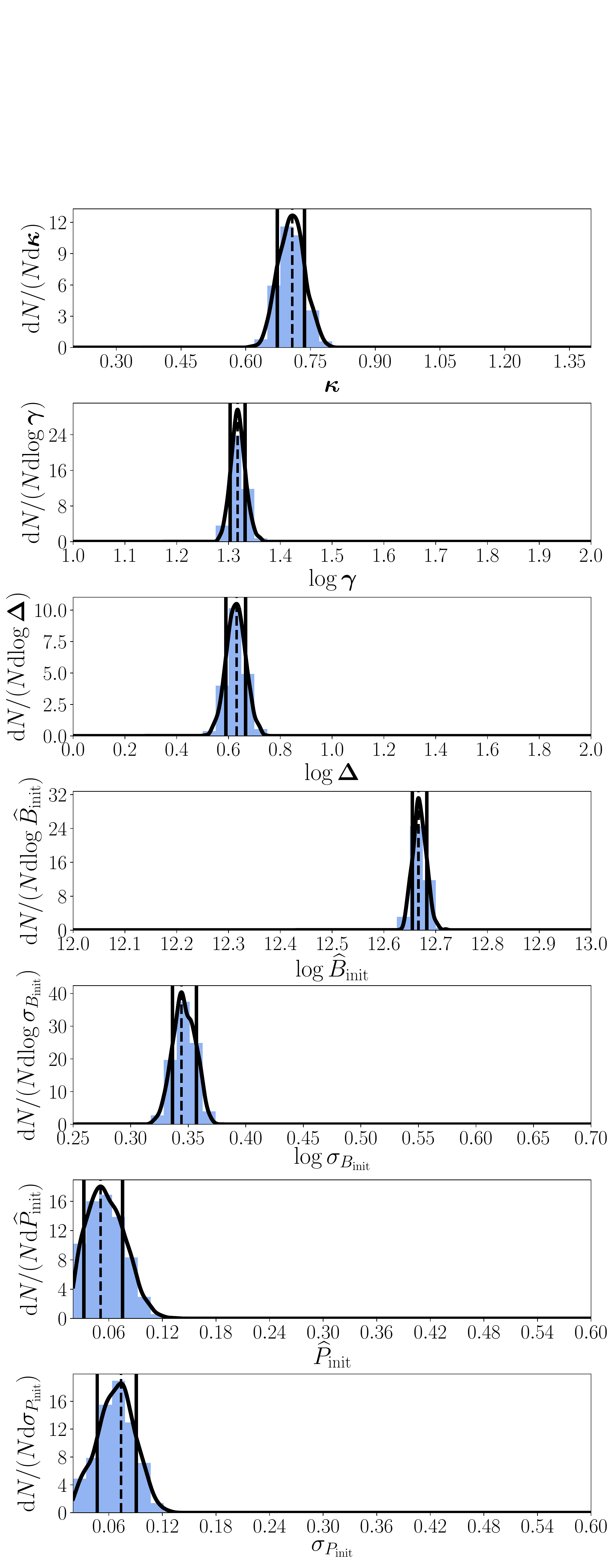} \\
\end{tabular}
\caption{The comparison of the marginal distributions for both models. The 1-sigma
    level is derived by using Gaussian Kernel Density Estimation (black  line) 
    by integrating the probability from the Most Probable Value (dotted line).
    We constrained the range of the $\gamma$ posterior to zoom in on the populated part of the phase space.} 
\label{fig:marginalall}
\end{figure*}

\subsection{The parameters marginal space}
To visualize the multidimensional parameter space, we present one and two dimensional
marginalised posterior distributions. The two dimensional results for
{\em power-law} and {\em rotational} models are presented in Figures 
\ref{fig:ContourPowerlLaw} and \ref{fig:ContourRotational}, respectively. The one-dimensional marginalised 
posterior distributions are shown in Figure \ref{fig:marginalall}.

\subsubsection{The marginal distribution}
The {\em two}-dimensional marginal probability distribution of the $i$-th and 
$j$-th parameters (denoted as $\mathcal{D}_{\theta_{ij}}$) is expressed by 
marginalizing the full-dimensional probability distribution $\mathcal{D}_{\bar{\theta}}$
upon all other parameters ($\Omega^{\prime}$ represents the parameter space excluding
$i$-th and $j$-th dimensions):
\begin{equation}
\mathcal{D}_{\theta_{ij}}=\int_{\Omega^{\prime}}\mathcal{D}_{\bar{\theta}}{\rm d}\Omega^{\prime}
\end{equation}
Similarly, the {\em one}-dimensional marginal distribution of the $i$-th parameter $\mathcal{D}_{\theta_{i}}$ is expressed by:
\begin{equation}
\mathcal{D}_{\theta_{i}}=\int_{\Omega^{\prime\prime}}\mathcal{D}_{\bar{\theta}}{\rm d}\Omega^{\prime\prime}
\end{equation}
where $\Omega^{\prime\prime}$ represents the parameter space excluding all but the $i$-th dimension.
To obtain the continuous probability density function ($\mathcal{C}_{\theta_{ij}}$
and $\mathcal{C}_{\theta_{i}}$) of the marginal distribution ($\mathcal{D}_{\theta_{ij}}$
and $\mathcal{D}_{\theta_{i}}$) we use the Gaussian kernel density estimation method 
\citep{scott2015multivariate} with a bandwidth $h$ (a function of number of points 
$n$ and dimensions $d$):
\begin{equation}
h(n,d) = n^\frac{-1}{d+4},
\end{equation}

\subsubsection{The most probable value and significance levels}
We denote the {\em most probable value} (MPV) -- the maximum of the marginal, 
continuous probability density function for each parameter (see Table \ref{tab:mvps}). For confidence levels 
we use the $1,2,3-\sigma$ ranges corresponding to the $68.27$, $95.45$, and $99.73$ 
per cent of the distribution. The $\sigma$ ranges are computed by integrating the 
probability around the MPV.
\begin{table}
\centering
\begin{tabular}{lr}
Parameter & Most Probable Value \\
\hline
    \multicolumn{2}{c}{{\em power-law} model} \\
\hline
    ${\alpha}$  &  $-0.37^{+0.22}_{-0.21}$ \\
    ${\beta}$ &  $0.26^{+0.04}_{-0.02}$ \\
    $\log {\left( \gamma/{\rm mJy}\right)}$ &  $1.40^{+0.06}_{-0.04}$ \\
    $\log {\left(\Delta/{\rm Myr}\right)}$  &  $0.56^{+0.10}_{-0.06}$ \\
    $\log \left( \widehat{B}_{{\rm init}}/{\rm G} \right)$ &  $12.66^{+0.01}_{-0.03}$ \\
    $\log \left( \sigma_{B_{{\rm init}}}/{\rm G} \right)$ &  $0.35^{+0.01}_{-0.01}$ \\
    $\widehat{P}_{{\rm init}}\,{\rm s}$ &  $0.03^{+0.03}_{-0.01}$ \\
    $\sigma_{P_{{\rm init}}}\,{\rm s}$  &  $0.04^{+0.04}_{-0.02}$ \\ 

\hline
    \multicolumn{2}{c}{{\em rotational} model} \\
\hline
    $\kappa$ &  $0.71^{+0.03}_{-0.04}$ \\ % & 0.63 & 0.82  \\
$\log {\left( \gamma/{\rm mJy}\right)}$  &  $1.32^{+0.01}_{-0.02}$ \\ %  & 1.25 & 1.4  \\
$\log {\left(\Delta/{\rm Myr}\right)}$  &  $0.63^{+0.04}_{-0.04}$ \\ %  & 0.55 & 0.78 \\
$\log \left( \widehat{B}_{{\rm init}}/{\rm G} \right)$ &  $12.67^{+0.01}_{-0.02}$ \\ %  & 12.63 & 12.7\\
$\log \left( \sigma_{B_{{\rm init}}}/{\rm G} \right)$  &  $0.34^{+0.02}_{-0.01}$ \\ %  & 0.33 & 0.38  \\
$\widehat{P}_{{\rm init}}\,{\rm s}$ &  $0.05^{+0.03}_{-0.02}$ \\ %  & 0.03 & 0.1\\
$\sigma_{P_{{\rm init}}}\,{\rm s}$  &  $0.07^{+0.02}_{-0.02}$ \\ %  & 0.03 & 0.11  \\

\hline
\end{tabular}
\caption{The most probable values (MPV) with one-$\sigma$ confidence level (the 
    upper and lower limit). The $\alpha$, $\beta$, and $\kappa$ parameters are
    dimensionless.}
\label{tab:mvps}
\end{table}

\subsubsection{Correlation coefficients}
In the Table \ref{tab:corr} we present the linear correlation coefficient $r$: 
\begin{equation}
r =\frac{\sum ^n _{i=1}(x_i - \bar{x})(y_i - \bar{y})}{\sqrt{\sum ^n _{i=1}(x_i - \bar{x})^2} \sqrt{\sum ^n _{i=1}(y_i - \bar{y})^2}}
\end{equation}
for both models for each pair of the parameters, where $n$ is the number of 
chains in the final analysis.
\begin{table}
\centering
\begin{tabular}{c}
The {\em power-law} model \\
\vspace{-1cm}
\includegraphics[width=\columnwidth]{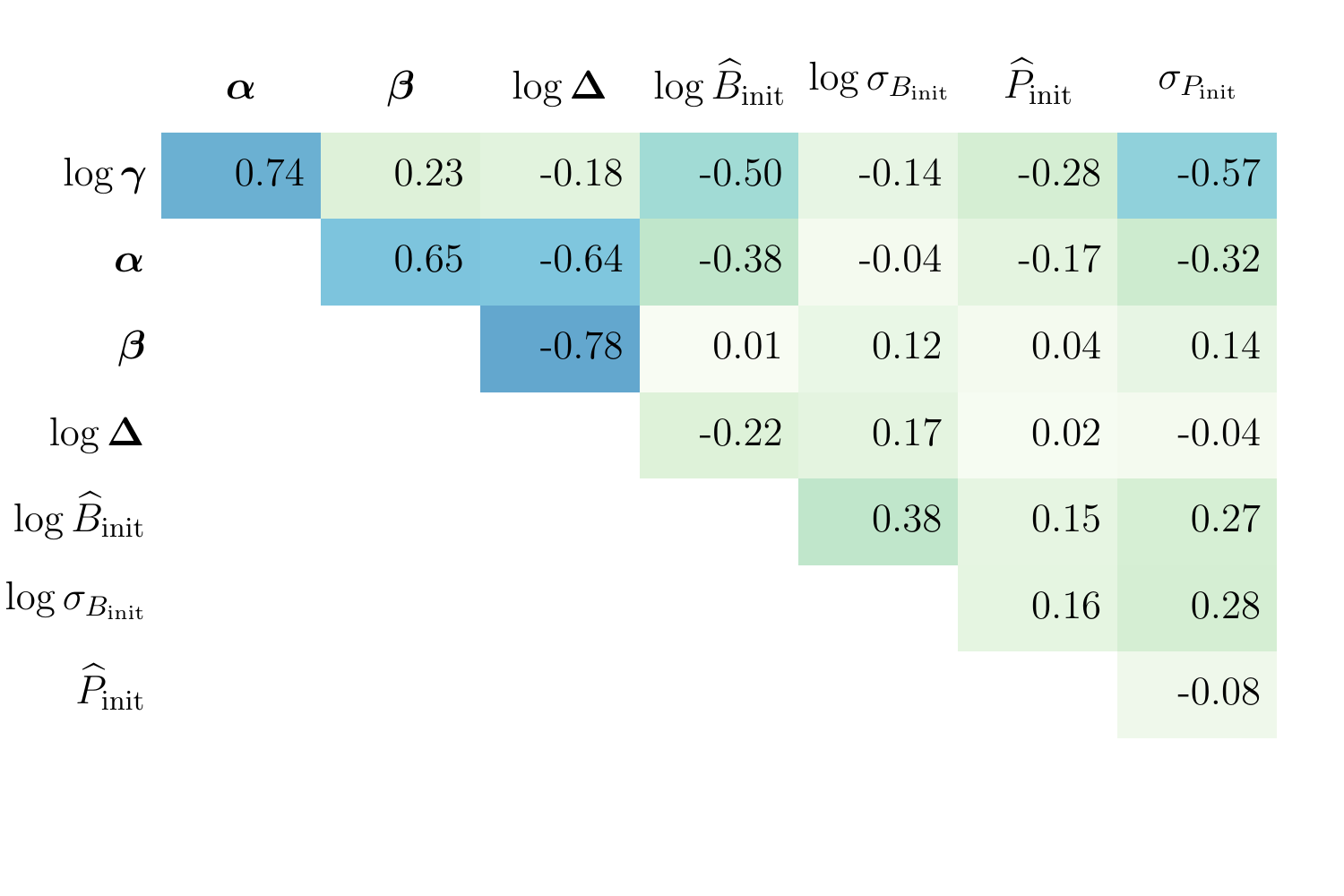} \\
The {\em rotational} model \\
\includegraphics[width=\columnwidth]{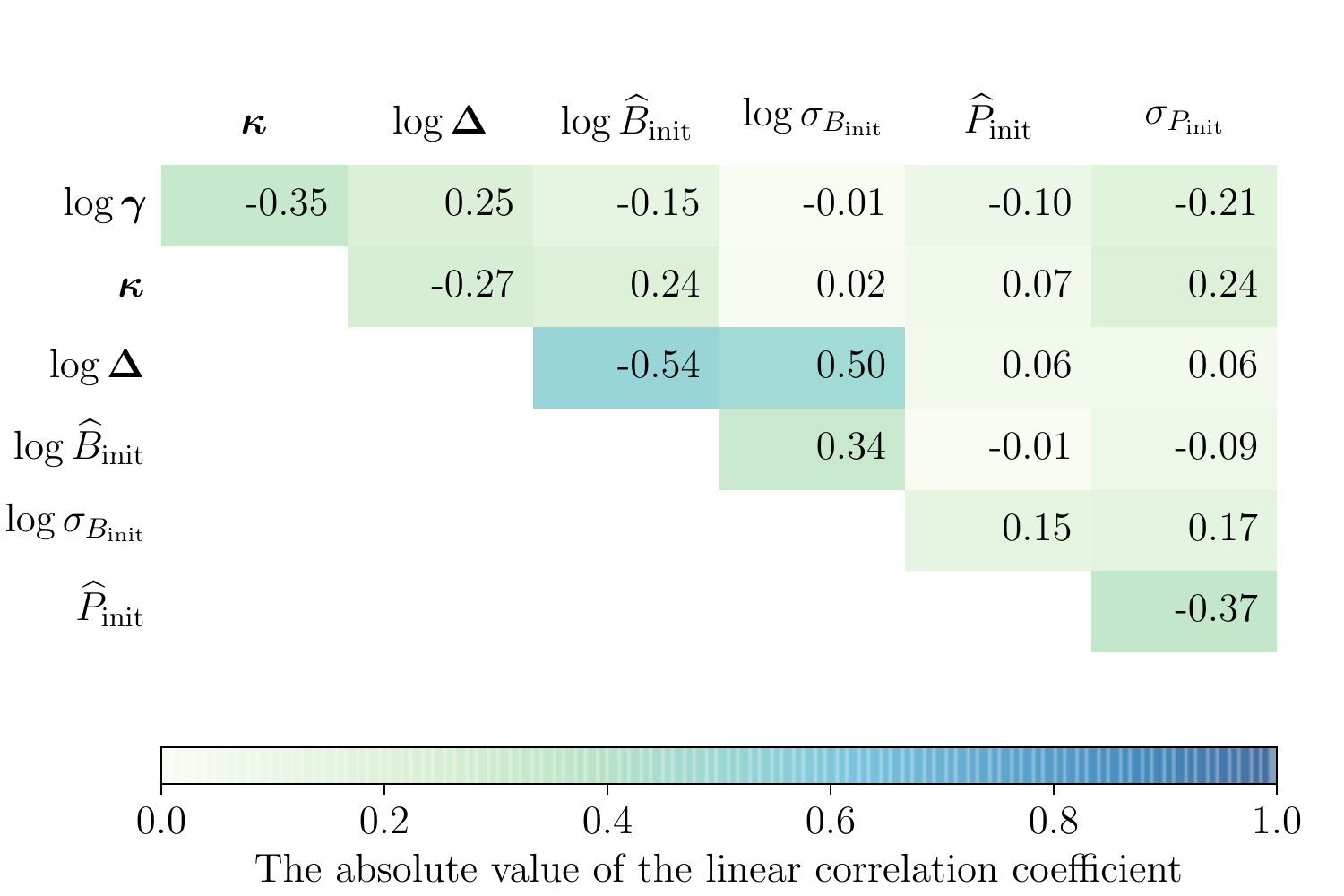} \\
\end{tabular}
\caption{The linear correlation coefficient matrices. The colour represent the 
    absolute value of correlation coefficients.} 
\label{tab:corr}
\end{table}

\subsection{The resulting population}
\begin{figure*}
    \begin{tabular}{c}
        ATNF Catalogue\\
        \includegraphics[width=\textwidth]{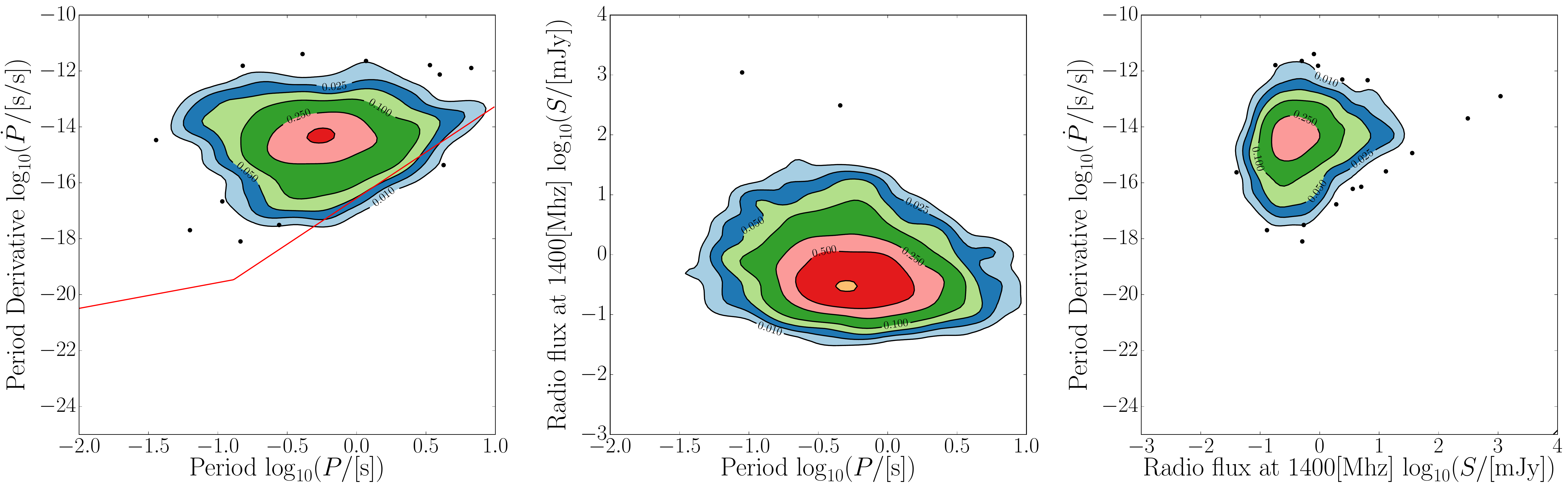} \\ 
        {\em Power-law} model\\
        \includegraphics[width=\textwidth]{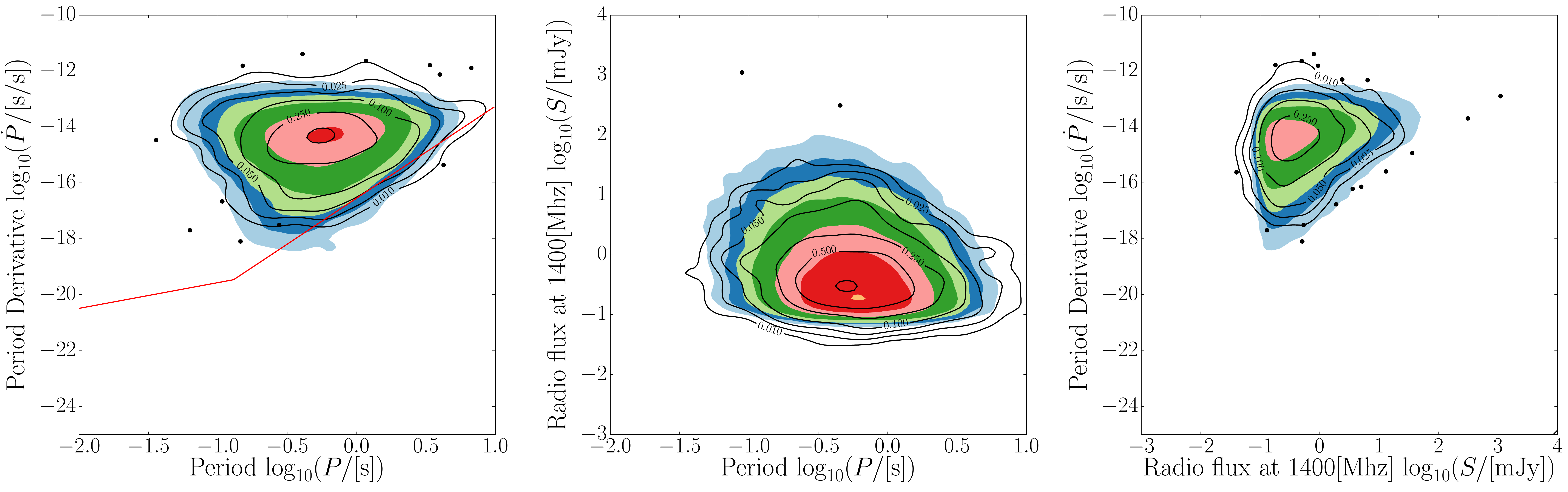} \\
        {\em Rotational} model\\
        \includegraphics[width=\textwidth]{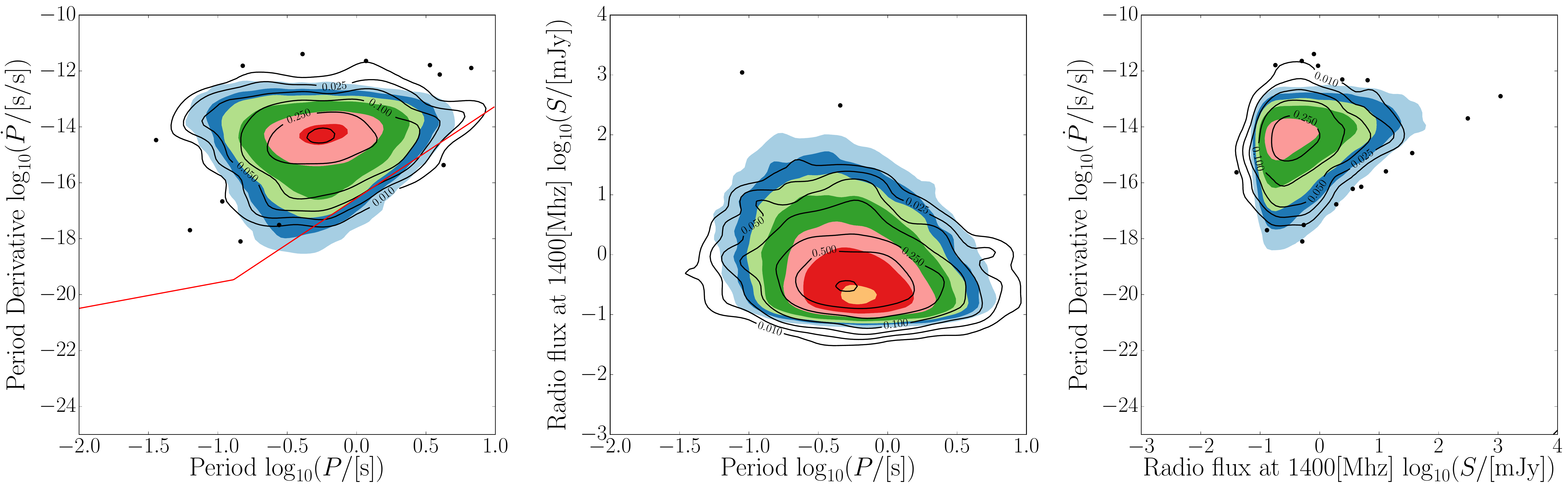} \\
        \includegraphics[width=\textwidth]{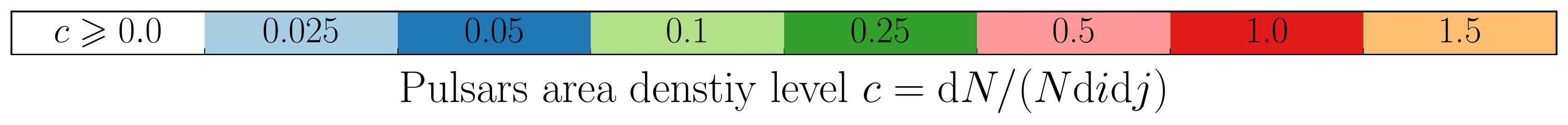}\\
    \end{tabular}
    \caption{Observations -- the subset of the ATNF catalogue (first row), 
    {\em power-law} model (second row), and 
    {\em rotational} model in the two-dimensional marginalisations of the comparison
    space $P$-$\dot{P}$-$S$. The colours indicate density levels for a given pair
    of dimensions $i$,$j$ (for e.g. $i=P$ and $j=\dot{P}$ for the first column).
    The contour lines relate to the observations in each plot. The dots represent 
    observations that lay in the region with extremely low density.}
    \label{fig:PopPPdotS}
\end{figure*}

For both sets of the MPVs (for the {\em power-law} and
{\em rotational} models) we computed a population of pulsars. We show the visible 
in the Parkes Multibeam Survey part of the population of the Figure \ref{fig:PopPPdotS}.
The method of presenting the pulsar density in two dimensional marginalisations 
of the comparison space ($P$-$\dot{P}$-$S$) is the final verification of the obtained
results. As can be seen in the second row for the {\em power-law} model and in the
third row of the Figure \ref{fig:PopPPdotS} for the {\em rotational} model, the fit
of the modelled data to the observations can not be considered incorrect. We note
that our simulation scheme always under estimates the data density -- this behaviour
can be seen as the pulsars density does not encompass the corresponding contour 
lines of the observations.

%%%%%%%%%%%%%%%%%%%%%%%%%%%%%%%%%%%%%%%%%%%%%%%%%%%%%%%%%%%%%%%%
%%%%%%%%%%%%%%%%%%%%%%%%%%%%%%%%%%%%%%%%%%%%%%%%%%%%%%%%%%%%%%%%
%%%%%%%%%%%%%%%%%%%%%%%%%%%%%%%%%%%%%%%%%%%%%%%%%%%%%%%%%%%%%%%%
%%%%%%%%%%%%%%%%%%%%%%%%%%%%%%%%%%%%%%%%%%%%%%%%%%%%%%%%%%%%%%%%
%%%%%%%%%%%%%%%%%%%%%%%%%%%%%%%%%%%%%%%%%%%%%%%%%%%%%%%%%%%%%%%%

\section{Discussion}\label{sec:discussion}
\subsection{The model}

\subsubsection{Initial period distribution}
We reached quite narrow initial period parameter distributions with most probable values equal to:
$\hat{P}_{\rm init}\approx 0.03\,{\rm s}$ and $0.05\,{\rm s}$ for the {\em power-law} and {\em rotational}
model, respectively. They both cover similar range of vales but the preferred 
mean value for the {\em power-law} model is significantly lower. 
In case of the standard deviation parameter of the initial period distribution  
we found out to be the most likely: $\sigma_{P_{\rm init}} \approx 0.03\,{\rm s}$ and $0.07\,{\rm s}$ 
for the  {\em power-law} and {\em rotational} model, respectively.

The resulting distributions of initial periods in both cases are in agreement with the predictions made by \citet{2007Natur.445...58B}.
However the hydrodynamic simulations lead to contradictory results \citep{2011ApJ...732...57R}.
In comparison to other population synthesis, in the works of \citet{2010MNRAS.401.2675P} 
they concluded that $\hat{P}_{\rm init}\approx 0.25\,{\rm s}$ with 
$\sigma_{P_{\rm init}}\approx 0.0001\,{\rm s}$, while \citet{2006ApJ...643..332F}
obtained the values $P_{\rm init}=0.3\,{\rm s}$ and $\sigma_{P_{\rm init}}=0.15\,{\rm s}$.
The main difference between our results and \citet{2006ApJ...643..332F}
is the inclusion of magnetic field decay which can be interpreted as an accelerator 
for the pulsar movement on the $P$-$\dot{P}$ plane.
Thus, the population can have faster initial periods as it
evolves to the same final population.
Moreover, \citet{2010MNRAS.401.2675P} included the magnetic field decay and reached similar
values as \citet{2006ApJ...643..332F}. 
Therefore, we are convinced the discrepancy with previous results is due to better sampling of the parameter space. In particular, we evaluated a larger number of models and didn't manually constrain the prior ranges.

\subsubsection{Initial magnetic field distribution}
The distribution of initial magnetic fields is almost 
identical in both models with the mean $\log B_{\rm init} \approx 12.66$ and
$\log \sigma_{B_{\rm init}} \approx 0.34$). Those results are consistent with
findings of \citet{2006ApJ...643..332F} where they obtained ${\log}B_{\rm init}=12.65$ 
and $\sigma_{B_{\rm init}}=0.55$. In the work of \citet{2010MNRAS.401.2675P},
authors reached larger value of the mean ${\log}B_{\rm init}=13.25$ with 
$\sigma_{B_{\rm init}}=0.6$.
Such initial distribution of values means that no pulsar has initial field less then
$ \log B_{\rm init} \approx 11$. Such conclusion  is consistent with the fact that
if such 
pulsars with low magnetic field strength existed they would be clearly observable in the radio band. Furthermore, their 
evolution would be very slow which would increase their detection probability. The lack of observed pulsars in the region of $P\approx 0.1\,{\rm s}$ and 
$\dot{P}\approx 10^{-17} - 10^{-18}\,{\rm ss}^{-1}$ implies that no quick spinning pulsars with magnetic field below  $ \log B_{\rm init} \approx 11$ are
formed.

\subsubsection{The {\em rotational} radio-emission model}
In the {\em rotational} model we obtained the value of the exponent $\kappa$ in 
range between $0.67$ and $0.74$. Upon translating to $L_{\rm rot}$ (see eq. \ref{eq:Lrot}),
we see that its exponent $\frac{1}{3}\kappa$ ranges from $0.22$ to $0.24$. This 
result disagrees with values obtained by \citet{2014MNRAS.443.1891G} in the range
from $0.45$ to $0.5$. We suspect that not including any radio switch-off mechanism 
(e.g. death lines) and limiting the comparison space to only $P$-$\dot{P}$ could 
play a significant role in the difference. 

\subsubsection{The {\em power-law} radio-emission model}
We found the {\em power-law} exponents to be in range from $0.24$ to $0.30$ 
for the $\beta$ and from $-0.58$ to $-0.15$ for the $\alpha$. 
We see that our most probable values, $\alpha=-0.58$ and $\beta=0.26$, are in
2-$\sigma$ range of the results obtained by \citet{2014MNRAS.439.2893B}: 
$\alpha = -1.12$ and $\beta = 0.28$. In comparison with \citep{2006ApJ...643..332F}: $\alpha=-1.5$, $\beta=0.5$,
we differ more than $3$ standard deviations. The difference can be explained by the inclusion of the magnetic field decay 
which does exhibit a strong correlation with other parameters describing the luminosity model
(see Table \ref{tab:corr} for $\alpha$, $\beta$, $\gamma$, and $\Delta$ parameters), and an improved sampling of the 
parameter space.

\subsubsection{The fit of radio luminosity laws}
\begin{figure}
    \includegraphics[width=\columnwidth]{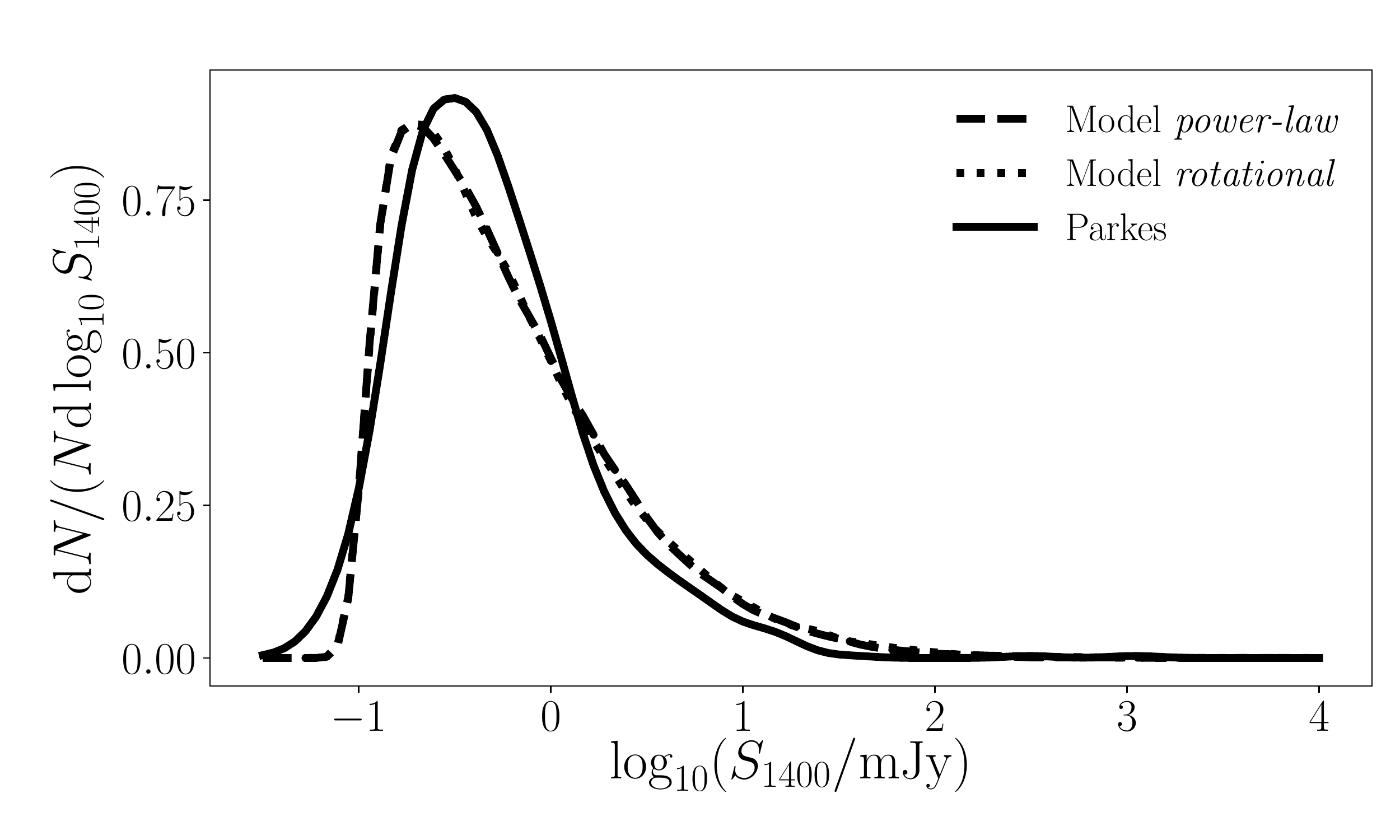}\\
    \caption{The comparison (upper plot) and difference (lower plot) of the distribution of the radio-flux of
    the observations, the {\em power-law} model, and the {\em rotational} model. }
    \label{fig:PowerLawHistMod}
\end{figure}

Although the fit in the 2D marginal distributions of the comparison space (second and third column of the Figure \ref{fig:PopPPdotS}) 
seems to be in general agreement with the observations, the comparison of the radio-flux distribution (see Figure \ref{fig:PowerLawHistMod})
shows some discrepancies. Our models underestimates the lower radio-fluxes, and 
overestimates the brightest objects. Both models behave in the same way pointing
to a possible systematic error in the method, or model description. 
The coupling of the $P$-$\dot{P}$-$S$ in both the optimisation (comparison space),
period evolution and radio-luminosity law, may have degenerated the problem --
leading to too few observational constraints with regard to the number of free parameters.
We also note that the introduction a phenomenological {\em death area} 
(see eq. \ref{eq:Death}) might have altered the distribution 
of pulsars in the $P$-$\dot{P}$ plane (see Figure \ref{fig:PopPPdotS}). We excluded the parameter
$\Psi$ from our current analysis due to the complexity reduction of the computations. 
We plan to include the analysis of the death area in our future work.

\subsubsection{The kicks distribution}
We are aware that \citet{2005MNRAS.360..974H} may be an imperfect distribution of 
the SN kicks, as stated in \citet{2006ApJ...643..332F} or more recently in 
\citet{2017A&A...608A..57V}. 
However, the spatial distribution (see section \ref{sec:performance} 
) is beyond the scope of this work. Moreover, by employing the Parkes Multibeam Survey, 
we focus only on the Galactic disk towards the centre of the Galaxy (see Table \ref{tab:ParkesMBParams}), thus limiting our
study to a younger subset of the whole Galactic population. Any possible discrepancies in 
the kicks model are neglected by this choice.

\subsubsection{Pulsar ages}
\label{sec:taudisc}
\begin{figure}
    \includegraphics[width=\columnwidth]{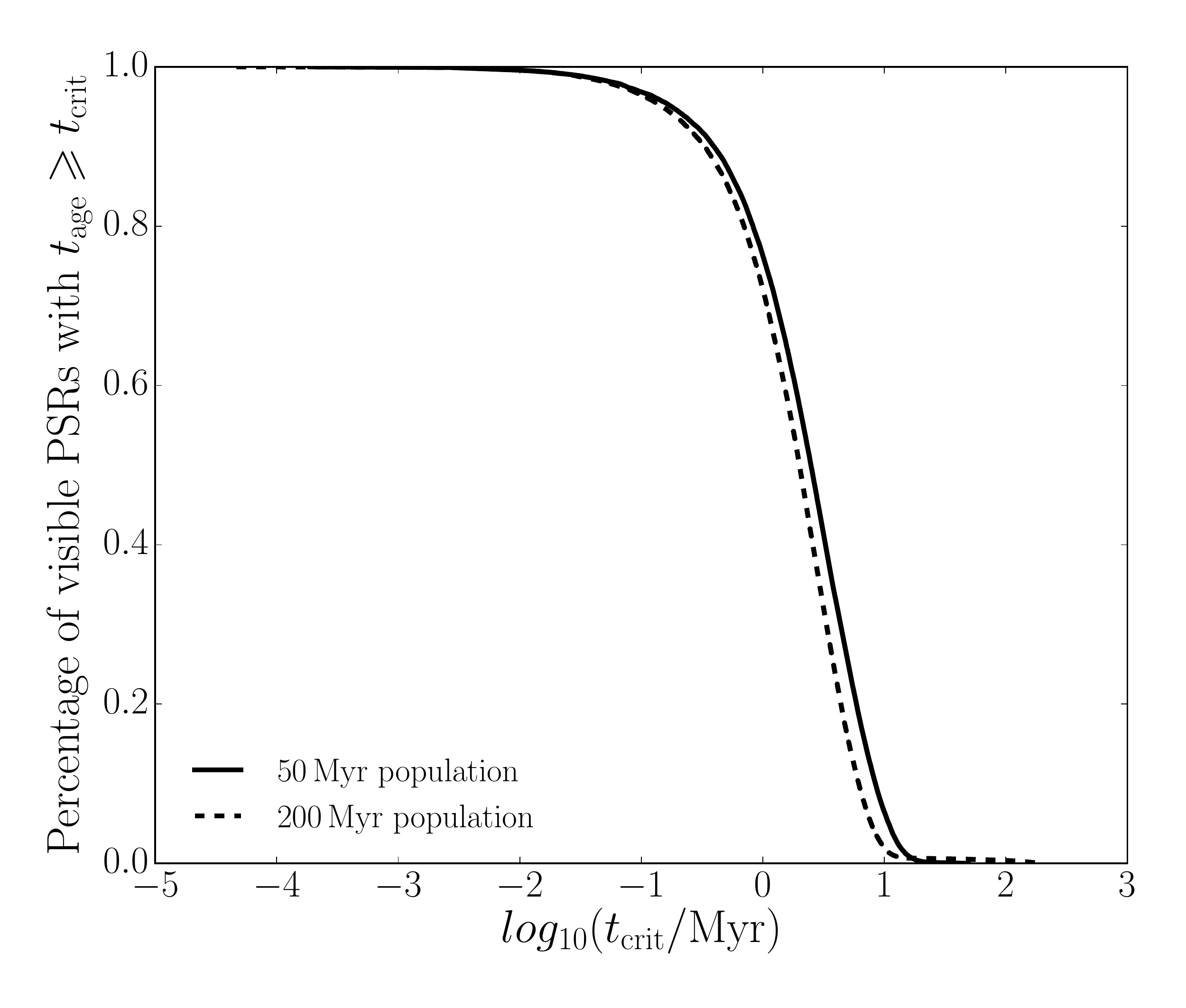}
    \caption{The cumulative distribution of the pulsars ages in the observed population. Lines -- simulation limited to maximum age $t_{\rm{age}}=50\,{\rm{Myr}}$, Dots -- simulation limited to  maximum age $t_{\rm{age}}=200\,{\rm{Myr}}$}.
    \label{fig:agecumul}
\end{figure}
\begin{figure}
\includegraphics[width=\columnwidth]{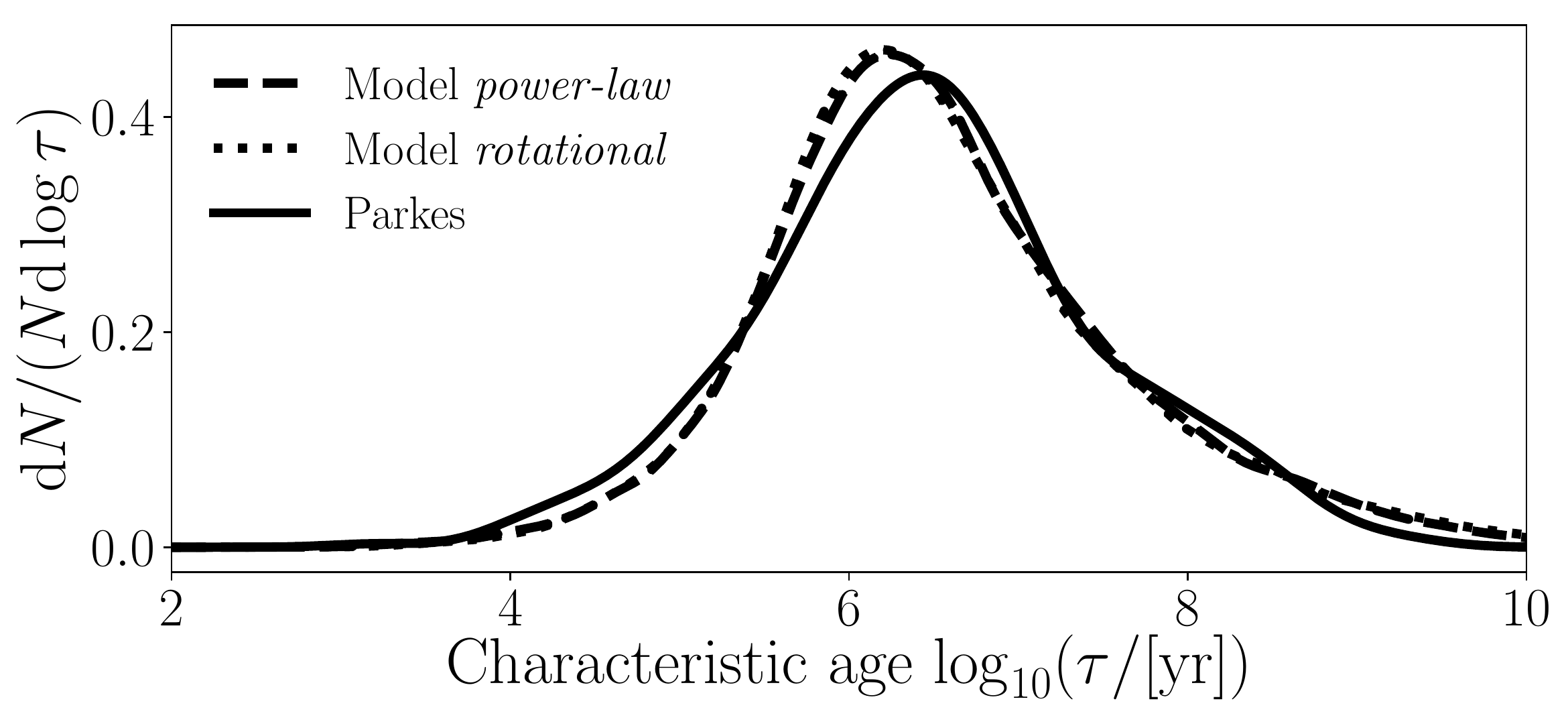} 
\caption{The probability density function of the characteristic age $\tau=P/(2\dot{P})$ in the reference {\em ATNF} subset for the Parkes Multibeam Survey and models.}
\label{fig:taudist}
\end{figure}

We found that, within our model, a fraction of less then $1\%$ of pulsars is older
then $50\,{\rm Myr}$ (see Figure \ref{fig:agecumul}). Because those pulsars do
not contribute in any significant way to the likelihood and as a result to MCMC and
parameter estimation (a smaller 200-chains test yielded similar results to presented therein), we neglected this part of population. We limited the maximum
age of pulsars to be $t_{\rm{age}}\leqslant50\,{\rm Myr}$. 
We do not contradict the observed kinematic ages distribution \citep{2013MNRAS.430.2281N}.
Our study is focused on the Galactic disk population (see Table \ref{tab:ParkesMBParams})
and we are unable to effectively compare with older kinematic-population. 
Moreover, by the inclusion of the magnetic field decay, we greatly speed up the
pulsars evolution track on the $P$-$\dot{P}$ plane. This may lead to incorrect comparison
between the evolution (simulation) age and the characteristic age distribution.
By computing the characteristic age
\begin{equation}
\tau=\frac{P}{2\dot{P}},
\end{equation}
we show in Figure \ref{fig:taudist} that the distribution of models resemble the 
observed sample and that it is possible to produce pulsars with $\tau{\geqslant}50\,{\rm Myr}$.
\begin{table}
\centering
\begin{tabular}{lccc}
Parameter & Symbol & Value & Unit \\
\hline
Receiver temperature & $T_{{\rm rec}}$ & $24$ & ${\rm K}$ \\
Bandwidth & $\Delta f$ & $288$ & ${\rm MHz}$ \\
Number of polarizations & $n_{{\rm p}}$ & $2$  & $1$\\ % 
Frequency & $f$ & $1400$ & ${\rm MHz}$ \\
Sampling time & $\tau_{{\rm samp}}$ & $0.00025$ & ${\rm s}$ \\
Gain & $G$  & $0.65$ & ${\rm K}/{\rm Jy}$ \\
Integration time & $t_{{\rm i}}$ & $2100$ & ${\rm s}$ \\
Diagonal dispersion measure & ${\rm DDM}$ & 27.61 & ${\rm pc}/{\rm cm}^3$ \\
System loss & $\iota$ & 1. & 1 \\
Min. signal to noise ratio & $\left({\rm S/N}\right)_{{\rm min}}$ & 10. & 1 \\
\hline
\end{tabular}
    The sky coverage 
\begin{tabular}{lc}
\hline
    Galactic longitude range & $-100^{\circ}\leqslant l \leqslant 50^{\circ}$ \\
    Galactic latitude range &  $-5^{\circ}\leqslant b \leqslant 5^{\circ}$ \\
\hline
\end{tabular}
\caption{Parkes Multibeam Survey parameters.}
\label{tab:ParkesMBParams}
\end{table}

\subsubsection{Estimated SN rate}
To derive the supernova rate from our models, we assume that the modelled and real populations have similar age ($t_{\rm max}=50\,{\rm Myr}$), and that the ratio of the number of visible pulsars $N_{\rm vis}$ to the total number of pulsars $N_{\rm tot}$ is constant:
\begin{equation}
    \frac{N_{\rm vis}^{\rm real}}{N_{\rm tot}^{\rm real}}=\frac{N_{\rm vis}^{\rm model}}{N_{\rm tot}^{\rm model}},
\end{equation}
where $N_{\rm vis}^{\rm model}$ is the number of visible pulsars in a given model equal to $60012$ and $58880$ for the {\em power-law} and {\em rotational} models respectively, the $N_{\rm tot}^{\rm model}$ is the total number of simulated pulsars equal to $5.e6$, the $N_{\rm vis}^{\rm real}$ is the observed sample of pulsars equal to $969$ pulsars, and the $N_{\rm tot}^{\rm real}$ is the total number of real pulsar and can be written as:
\begin{equation}
    N_{\rm tot}^{\rm real}=t_{\rm max} r_{\rm SN}.
\end{equation}
Where the $r_{SN}$ is the supernova rate per century. Thus, we obtain our estimate:
\begin{equation}
    r_{\rm SN}=\frac{ N_{\rm vis}^{\rm real} N_{\rm tot}^{\rm model}  }{ N_{\rm vis}^{\rm model} t_{\rm max}  },
\end{equation}
which yields:
\begin{equation}
    r_{\rm SN}=
    \begin{cases} 
        0.1615/100\,{\rm yr} & \text{{\em power-law} model}  \\
        0.1646/100\,{\rm yr} & \text{{\em rotational} model}
    \end{cases}
\end{equation}
Our estimate do not exceed the predicted rate of core-collapse supernova rate $0.5-1.1$ per century \citep{1991ARA&A..29..363V} as well the recent estimate based on INTEGRAL data for the combined type I b/c and type II supernova rate equal to $1.9\pm1.1$ per century \citep{2006Natur.439...45D}.

\subsubsection{The DM distribution } 
\begin{figure}
    \includegraphics[width=\columnwidth]{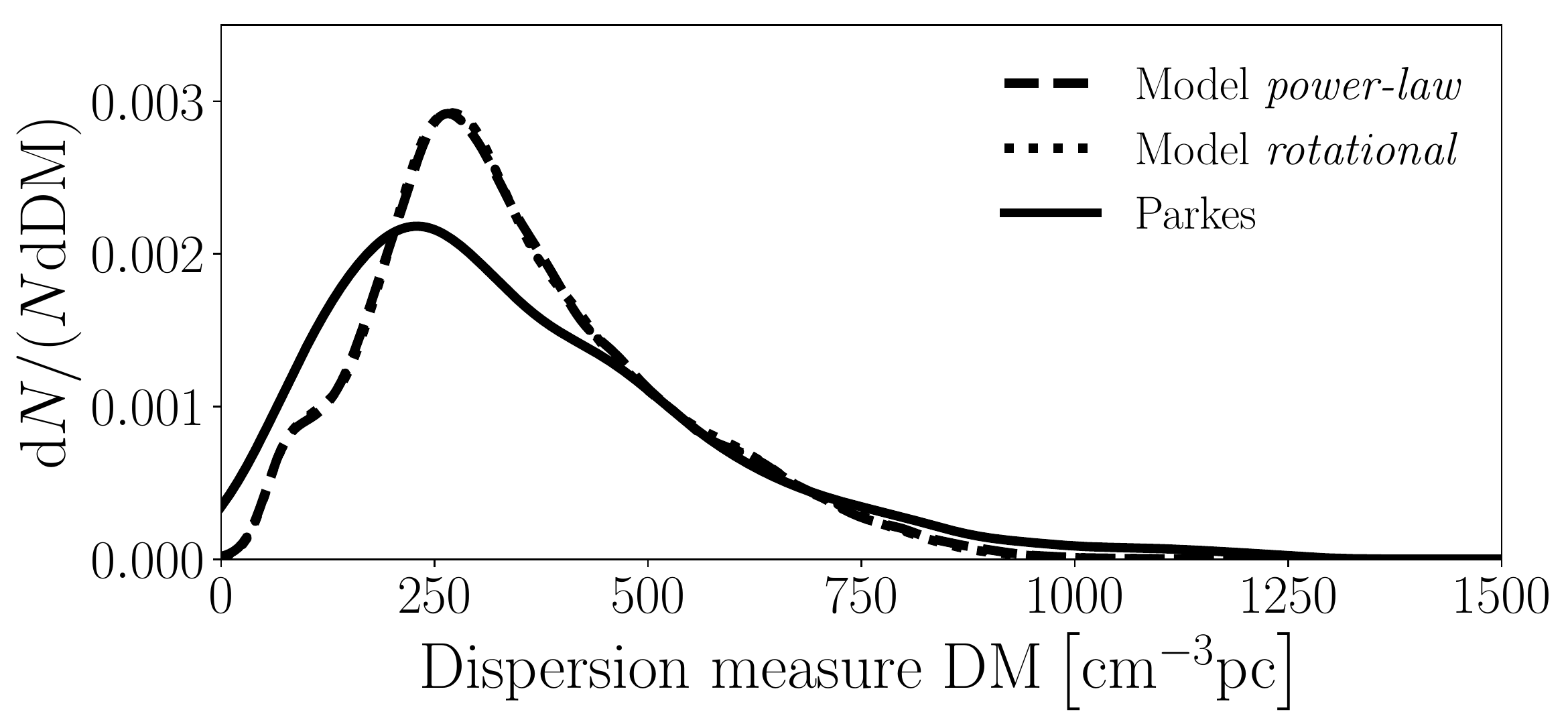}
    \caption{The probability density function of the dispersion measure in the 
    reference {\em ATNF} subset for the Parkes Multibeam Survey and models.}
    \label{fig:DMdist}
\end{figure}

We limited our comparison space to three dimensions only -- the period $P$ and 
its derivative $\dot{P}$ and the radio flux $S_{1400}$. Due to restriction on the 
computational time of the free electron distribution model of 
\citet{2017ApJ...835...29Y} we excluded the geometrical part (galactic coordinates 
and dispersion measure) from our comparison space. Therefore we do not draw any 
conclusion about the pulsars spatial distribution in the Milky Way and their 
initial kicks, but the distances (in our case the dispersion measure) can have 
implication for the radio luminosity model. We present the model distribution of 
the dispersion measure and the one of the Parkes Multibeam Survey in Figure 
\ref{fig:DMdist}. The model distribution and the observed one are close even 
though they were not fitted.

\subsection{Pulsar population with {\em Square Kilometre Array}}
\begin{figure}
\centering
\begin{tabular}{c}
The {\em power-law} model \\
\includegraphics[width=\columnwidth]{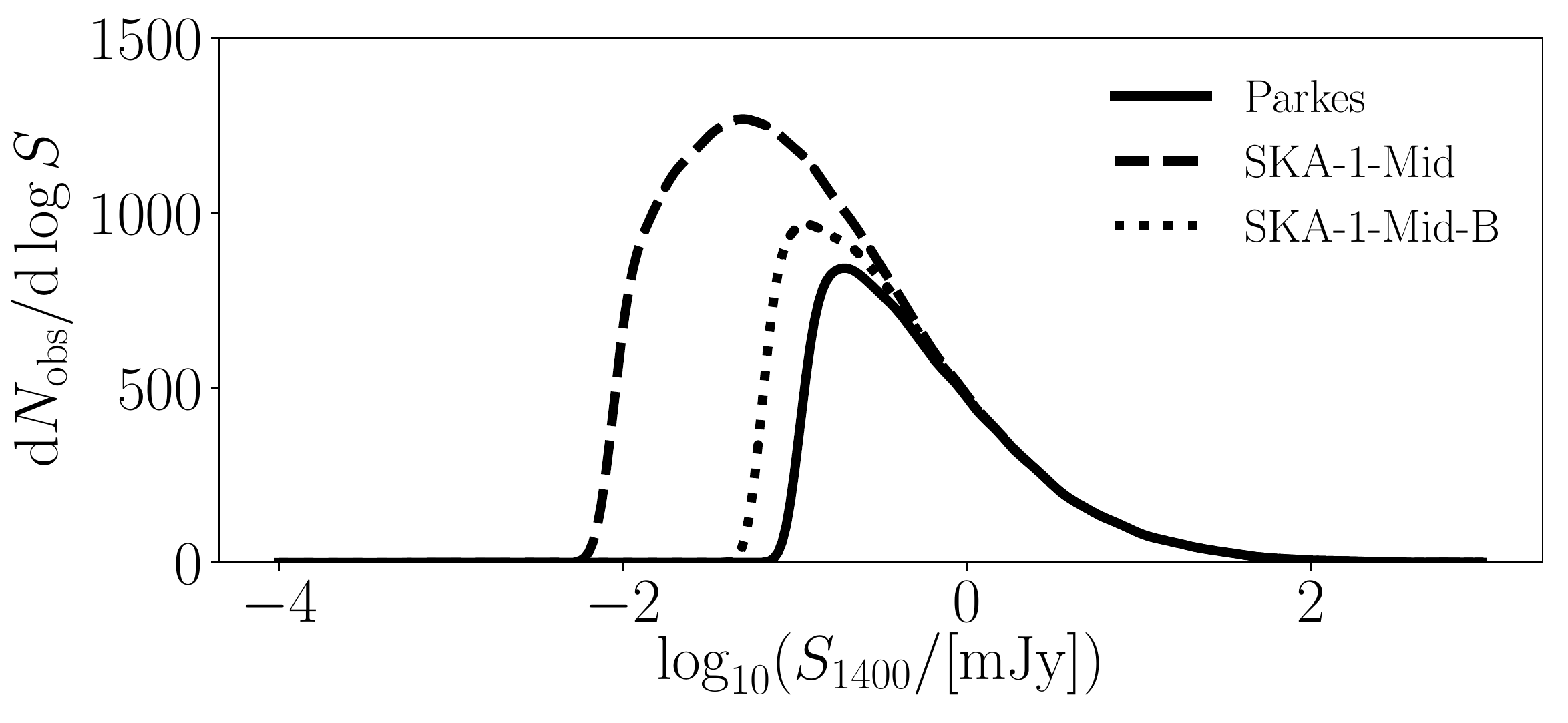} \\
The {\em rotational} model \\
\includegraphics[width=\columnwidth]{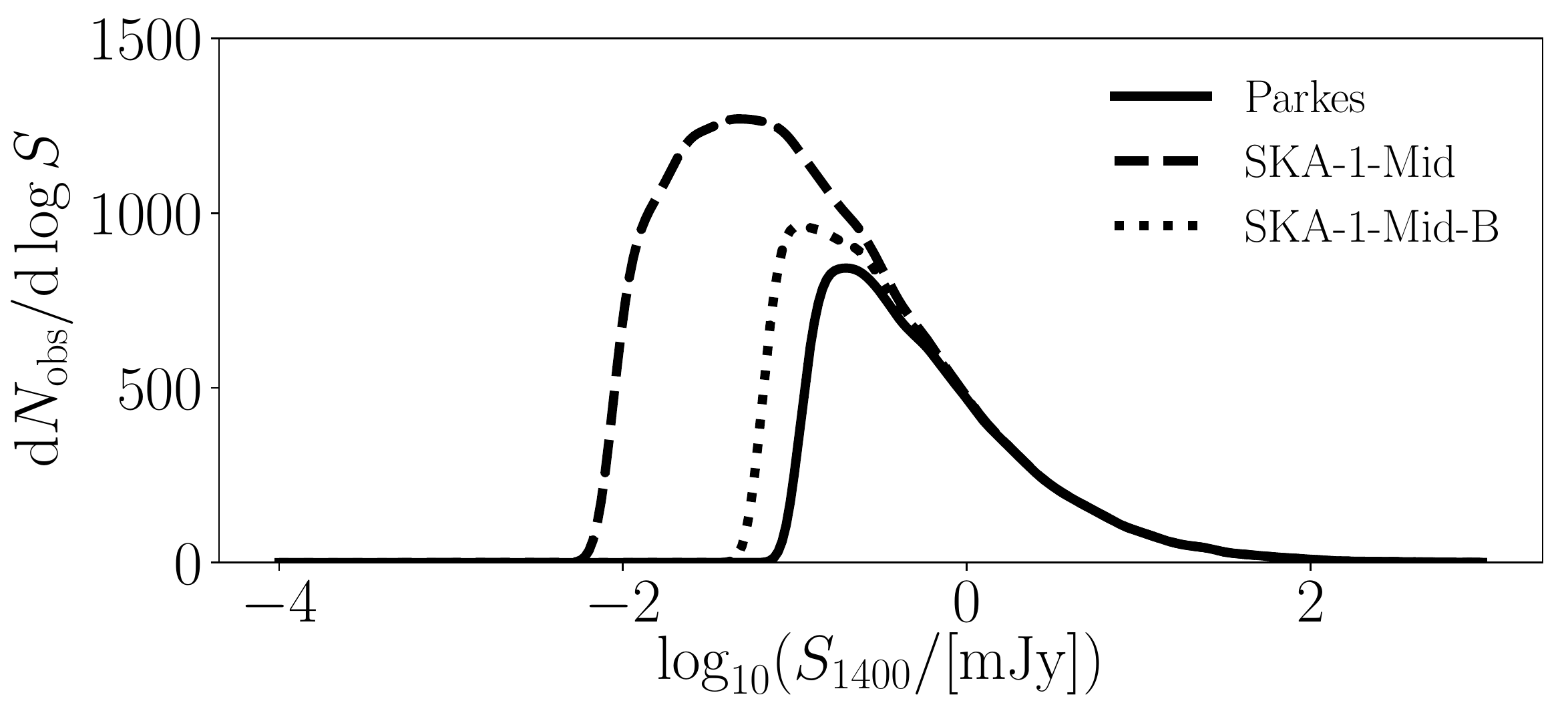} \\
\end{tabular}
\caption{The distribution of pulsars radio-fluxes in the Parkes Multibeam Survey 
    and the prediction for the SKA-1-Mid and SKA-1-Mid-B surveys. The total number of pulsars in $N_{\rm obs}$ is equal to $969$ for Parkes with our selection cuts. In case of the SKA surveys the number of pulsars is $2364$ and $2285$ for the {\em power-law} and {\em rotational} model, respectively. For the {\em low-cost SKA} survey the number of pulsar is almost the same $1241$ and $1229$ for the {\em power-law} and {\em rotational} model, respectively.}
\label{fig:SKAS1400} 
\end{figure}

A very interesting consequence of the modelling presented here is the possibility
to extend the results to the population of pulsars observable by the {\em Square 
Kilometre Array} (SKA). The SKA telescope is described in \citet{2004NewAR..48..979C},
\citet{ska2015advancing}, \citet{2015aska.confE..36K}, and \citet{2017ARep...61..288G}.
We present the extrapolation of the observable pulsar population using the two 
models of pulsar luminosity considered above (the {\em power-law} model and 
{\em rotational} one) with their parameters set to the {\em most probable values} 
(see Table \ref{tab:mvps}). We list two sets of probable parameters that describe the SKA 
for a mid-frequency survey in Table \ref{tab:SKAParams}. The first one SKA-1-Mid 
represents our estimate of the initially planned SKA operation and the SKA-1-Mid-B
a more pragmatic view of the parameters.
\begin{table}
\centering
\begin{tabular}{lccc}
Parameter & Symbol & Value & Unit \\
\hline
Receiver temperature & $T_{{\rm rec}}$ & $30$ & ${\rm K}$ \\
Bandwidth & $\Delta f$ & $300$ & ${\rm MHz}$ \\
Number of polarizations & $n_{{\rm p}}$ & $2$ & $1$\\ 
Frequency & $f$ & $1400$ & ${\rm MHz}$ \\
Sampling time & $\tau_{{\rm sampling}}$ & $0.000064$ & ${\rm s}$ \\
Gain & $G$  & $15$ ($2$) & ${\rm K}/{\rm Jy}$ \\
Integration time & $t_{{\rm i}}$ & $2100$ & ${\rm s}$ \\
Diagonal dispersion measure & ${\rm DDM}$ & 289.49 & ${\rm pc}/{\rm cm}^3$ \\
System loss & $\beta$ & 1. & 1 \\
Min. signal to noise ratio & $\left({\rm S/N}\right)_{{\rm min}}$ & 10. & 1 \\
\hline
\end{tabular}
\caption{SKA-1-Mid and SKA-1-Mid-B (in brackets) survey parameters.}
\label{tab:SKAParams}
\end{table}
To perform the extrapolation we compute a pulsar population of a given size ($10^{7}$)
for the best set of parameters for both models.
We infer what part of this population is seen in each survey (Parkes Multibeam,
SKA-1-Mid, and SKA-1-Mid-B). We then compare the ratio of modelled pulsars
seen in the Parkes Multibeam Survey to the cardinality of used subsection of the
ATNF catalogue. This ratio is considered the normalisation constant $W$. In order
to scale the artificial SKA observation we restrict the SKA-1-Mid and SKA-1-Mid-B
surveys to the same part of the sky as the Parkes Multibeam Survey. Upon scaling
the SKA surveys with the normalisation constant $W$ we reach the estimated number
of detectable single pulsars. We present the distribution of the detectable pulsars
in the function of the radio-flux at Figure \ref{fig:SKAS1400}. By our estimate,
should the SKA observatory perform a survey of the same part of sky as the Parkes
Mutlibeam Survey, we would reach an increase in detected radio pulsars by $23\%$ or
$137\%$ for the SKA-1-Mid-B and SKA-1-Mid survey, respectively.

\section{Conclusions}\label{sec:conclusions}
We presented a radio pulsar population synthesis model based on the one described by \citet{2006ApJ...643..332F}. We compared it with observations using the likelihood statistic 
and we used the Markov Chain Monte Carlo method to explore the parameter space.
We used the recent model for the computation of the interstellar medium by 
\citet{2017ApJ...835...29Y}. We compared the model with observations in the space spanned
by period, period derivative, and radio flux. The pulsar initial parameters and their 
evolution are described by five parameters. We have explored two models with two different 
parametrizations of the pulsar luminosity: the {\em power-law} model described by three 
additional parameters and the {\em rotational} model described by two parameters. 
This allowed us to the estimation of the parameters and their confidence level.
We found that the magnetic field decay is scale quite short 
and is approximately $3.63^{+0.94}_{-0.47}\,{\rm Myr}$ in the {\em power-law} or $4.27^{+0.4}_{-0.38}\,{\rm Myr}$ in the {\em rotational} model. 
The initial period distributions are centred around $\hat{P}_{\rm init}\approx 30\,{\rm ms}$ and $50\,{\rm ms}$ 
with widths of $\sigma_{P_{\rm init}}\approx 40\,{\rm ms}$ and $70$ for the {\em power-law} and {\em rotational} model, respectively.
The initial distribution of logarithm of magnetic field is almost identical in both models with the average 
$\log B_{\rm init} \approx 12.66$ and $\log \sigma_{B_{\rm init}} \approx 0.35$.
We found that the preferred values of the exponents for the {\em power-law} radio-luminosity 
model are $\alpha=-0.37^{+0.22}_{-0.21}$ and $\beta=0.26^{+0.04}_{-0.02}$, and for the model 
proportional to the rotational energy loss is $\kappa=0.71^{+0.03}_{-0.04}$.
Proposed parameters values differ from the works of \citet{2006ApJ...643..332F},
\citet{2010MNRAS.401.2675P}, \citet{2014MNRAS.443.1891G}, and \citet{2014MNRAS.439.2893B}.
We contribute the difference to the inclusion of the radio-flux in the space that
the main statistic (used to optimize the model) used as well as significantly 
large scope of the simulations (due to the increase of the available computational
power). 

In the view of the significant linear correlation between parameters exhibited by the
{\em power-law} model, and the fact that two models lead to an almost identical population of observed pulsars, 
we believe that the {\em rotational} model should be preferred. We note, that even the {\em rotational} model
has some non-negligible correlation between magnetic field decay time-scale and 
parameters describing the initial distribution of magnetic field. To shed some light on the possible cause of the 
parameters correlation, additional constraints from the observation should be provided, and more sophisticated description of the radio luminosity implemented. 

We estimated the number of new observable pulsars, should the SKA survey cover the 
same area, as the Parkes Multibeam Survey to be increased by $23-137\%$ depending
on the final parameters of the SKA survey.
We release the {\em Indri} code\footnote{\url{http://github.com/cieslar/Indri}}
used in this research in hopes of contributing to the advancement of {\em dynamical} 
models in the pulsar population synthesis research field.
%%%%%%%%%%%%%%%%%%%%%%%%%%%%%%%%%%%%%%%%%%%%%%%%%%

\section*{Acknowledgements}
Marek Cie\'{s}lar and Tomasz Bulik were supported by the NCN Grant No. 
UMO-2014/14/M/ST9/00707. Marek Cie\'{s}lar acknowledges support from Polish 
Science Foundation {\em Master2013} Subsidy as well as from the NCN Grant No. 2016/22/E/ST9/00037. 
Tomasz Bulik is grateful for support from TEAM/2016-3/19 from the FNP.
Stefan Os{\l}owski acknowledges support from the Alexander von Humboldt Foundation and ARC grant Laureate Fellowship FL150100148.\\
In our work we used the Mersenne Twister pseudo random number generator by
\citet{Matsumoto:1998:MTE:272991.272995}. We thank Micha{\l} Bejger and Pawe{\l} 
Cieciel\c{a}g from Nicolaus Copernicus Astronomical Center 
(Polish Academy of Sciences, Warsaw) for housing part of our simulations on the 
{\em bigdog} cluster located at the Institute of Mathematics (Polish Academy of 
Sciences, Warsaw).
The majority of the computations were performed on the OzSTAR national facility at Swinburne University of Technology. OzSTAR is funded by Swinburne University of Technology and the National Collaborative Research Infrastructure Strategy (NCRIS).

\bibliographystyle{mnras}
\bibliography{Radio} % if your bibtex file is called example.bib

%%%%%%%%%%%%%%%%%%%%%%%%%%%%%%%%%%%%%%%%%%%%%%%%%%
%%%%%%%%%%%%%%%%% APPENDICES %%%%%%%%%%%%%%%%%%%%%
\appendix
%%%%%%%%%%%%%%%%%%%%%%%%%%%%%%%%%%%%%%%%%%%%%%%%%%
\section{}
\label{sec:FK06reimplementation}
In this appendix, for the completion purpose, we present the parts of the model
that are identical to the model developed by \citet{2006ApJ...643..332F}. 

\subsection{The Milky Way}
\subsubsection{The Galactic potential}
We use the well-established three-component Galactic potential consisting of the 
disk, the bulge and the halo.
The bulge $\Phi_{{\rm {Bulge}}}$ and the disk $\Phi_{{\rm {Disk}}}$ 
gravitational potentials are adopted after \citet{1975PASJ...27..533M}. The 
formula describing the bulge is:
\begin{equation}
\Phi_{{\rm {Bulge}}} = -\frac{G M_{b}}{\sqrt{b_{b}^{2} + r^{2}}}
\end{equation} 
where $M_b=1.12\times 10^{10}\,{\rm M}_{\odot}$ and $b_b=0.277\,{\rm kpc}$, 
and $r=(x^2+y^2+z^2)^{1/2}$.  We model the disk potential as: 
\begin{equation}
\Phi_{{\rm {Disk}}} = -\frac{G M_{d}}{\sqrt{{\left(a_{d} + \sqrt{b_{d}^{2} + z^{2}}\right)}^{2} + \rho^{2}}}
\end{equation}
where $M_d=8.78\times 10^{10}\,{\rm M}_{\odot}$, $a_d=4.2\,{\rm kpc}$ 
and $b_d=0.198\,{\rm kpc}$, and 
$\rho=(x^2+y^2)^{1/2}$.
We use the halo potential $\Phi_{{\rm {Halo}}}$ following the model of \citet{1990ApJ...348..485P}:
\begin{equation}
\Phi_{{\rm {Halo}}} = -\frac{G M_{h}}{2 r_{c}} {\left(2\frac{ r_{c}}{r} \arctan\left(\frac{r}{r_{c}}\right) + \log\left(\frac{r^{2}}{r_{c}^{2}} + 1\right)\right)}
\label{eq:PaczynskiHalo}
\end{equation}
where $M_h=5\times 10^{10}\,{\rm M}_{\odot}$ $r_c=6\,{\rm kpc}$. As the 
associated density of the halo is diverging so 
we cut the halo potential at $r_{cut}=100\,{\rm kpc}$, see e.g. \citet{2010ApJ...725..816B}.
We neglect the dependence of the galactic potential on the individual Galactic arms.
%%%%%%%%%%%%%%%%%%%%%%%%%%%%%%%%%%%%%%%%%%%%%%%%%%
\subsubsection{The initial positions of pulsars}
\begin{figure}
	\includegraphics[width=\columnwidth]{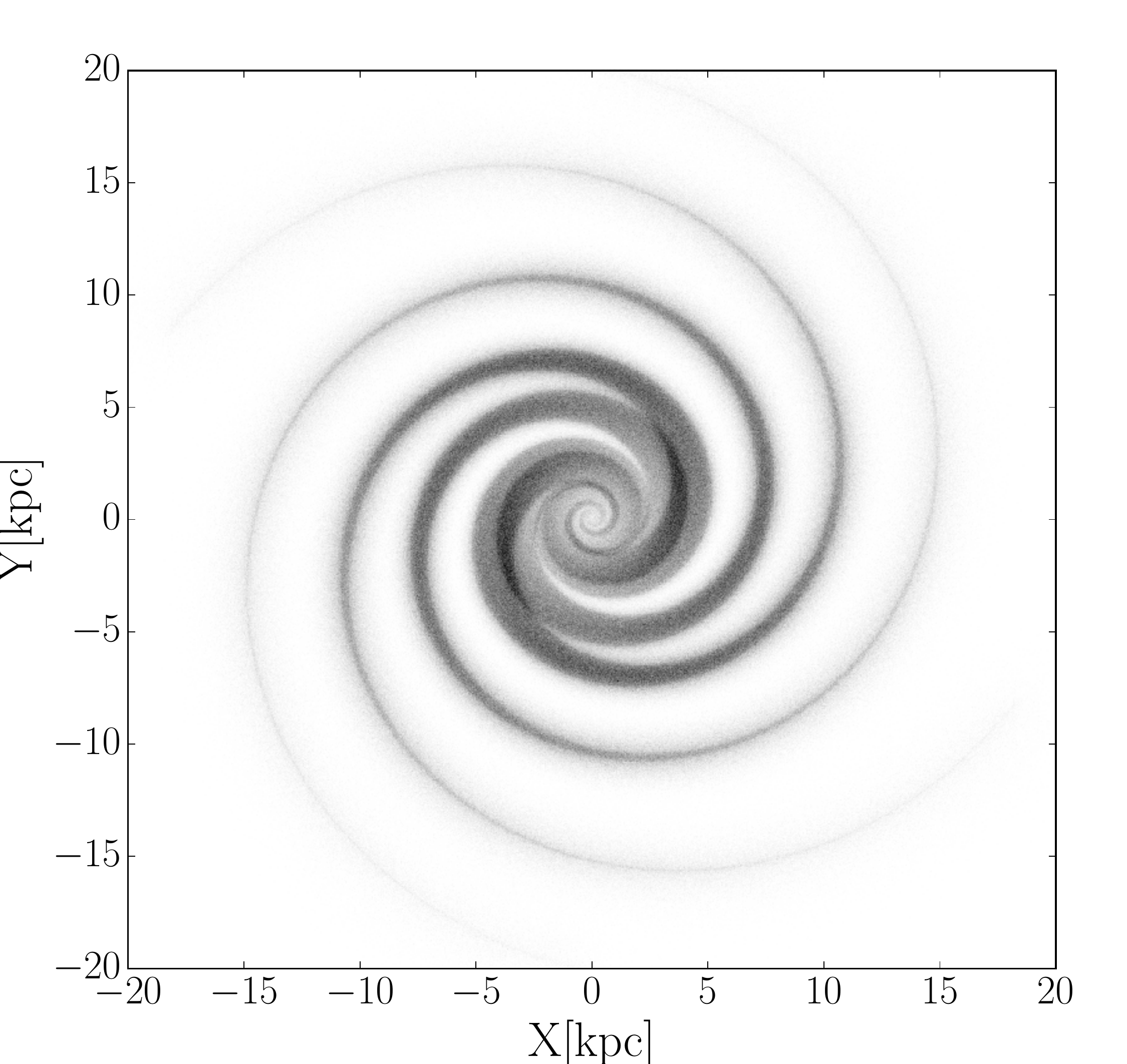}
	\caption{Spiral Arms -- initial positions distribution in the Galactic plane 
    of randomly drawn, $10^{7}$ pulsars.}
	\label{fig:PulsarsInitPosition}
\end{figure}

\begin{table}
\centering
\begin{tabular}{lccc}
Arm & $k\,{\rm rad}$ & $\rho_{0}\,{\rm kpc}$ & $\theta_{0}\,{\rm rad}$ \\
\hline
Norma & $4.25$ & $3.48$ & $1.57$ \\
Carina-Sagittarius & $4.25$ & $3.48$ & $4.71$ \\
Perseus & $4.89$ & $4.90$ & $4.05$ \\
Crux-Scutum & $4.89$ & $4.90$ & $0.95$ \\
\hline
\end{tabular}
\caption{Spiral arms parameters.}
\label{tab:SpiralArmsParameters}
\end{table}
We adopt the initial position distribution after \citet{2006ApJ...643..332F} with 
the assumption that pulsars are born inside the galactic spiral arms. Following them, 
we exclude the Local Arm as the origin of the pulsars. 
The centroids of each arm  are described as  logarithmic spirals \citep{1992ApJS...83..111W}: 
\begin{equation}
\theta(\rho)=k\log\left(\frac{\rho}{\rho_0}\right)+\theta_{0}
\label{equation:SpiralArm}
\end{equation}
with their parameters listed in the Table \ref{tab:SpiralArmsParameters}. With 
equal probability we chose the arm in which pulsar is born.
The distance $\rho_{{\rm raw}}$ from the centre of the Galaxy is drawn using 
the stellar surface density distribution in the Galactic plane \citep{2004A&A...422..545Y}:
\begin{equation}
\xi(\rho_{{\rm raw}}) \sim \left(\frac{\rho_{{\rm raw}} + R_1}{R_{\odot} + R_1} \right)^{a} \exp \left(-b\left(\frac{\rho_{{\rm raw}}-R_{\odot}}{R_{\odot}+R_1}\right)\right)
\end{equation}
where $a=1.64$, $b=4.01$, $R_1=0.55\,{\rm kpc}$, and $R_\odot=8.5\,{\rm kpc}$ 
is the distance of the Sun from the Galactic centre.
We insert the radial distance into equation \ref{equation:SpiralArm} to obtain the 
position along the spiral arm's centroid $(\rho_{{\rm raw}},\theta_{{\rm raw}})$. 
This position is then smeared by adding a correction to the angle $\theta_{{\rm raw}}$ 
to avoid artificial structures in the Galactic centre:
\begin{equation}
\theta_{{\rm wide}} = \theta_{{\rm raw}} + \theta_{{\rm corr}}\exp{\frac{-0.35\rho_{{\rm raw}}}{{\rm kpc}}}
\end{equation}
where $\theta_{{\rm corr}}$ is randomly chosen from the interval of $[0,2\pi)$ radians.
We introduce the internal structure of spiral arms by displacing the initial 
radial position of the pulsar in the galactic plane. We add a vector with random
direction and a length drawn from a Gaussian distribution with 
$\sigma=0.07\rho_{{\rm raw}}$. The resulting initial position distribution in 
the Galactic plane is shown in the Figure \ref{fig:PulsarsInitPosition}.
The vertical position of the pulsar is drawn from the exponential distribution 
with the mean $\left<z_{0}\right>=0.05\,{\rm kpc}$.
We populate the Galaxy with stars by rotating spiral arms and inserting pulsars 
uniformly in time from their maximal simulated age, 
$\max(t_{\rm age})=50\,{\rm Myr}$ ago, to a present day. We assume a simple, 
rigid Galactic rotation with the period of $P_{\rm rot} =250\,{\rm Myr}$: 
\begin{equation}
\theta = \theta_{{\rm wide}} - 2\pi t_{\rm age}/P_{\rm rot}
\end{equation}
We justify the rigid rotation assumption with the maximum possible age of a 
modelled pulsar $\max(t_{\rm age})=50\,{\rm Myr}$ being significantly lower 
then the rotation period of the Galaxy. For the discussion of this assumption see 
section \ref{sec:taudisc}.

\subsubsection{The initial velocity}
At birth, each pulsar is subjugated to a kick due to the supernova explosion
resulting in change in the initial velocity. We use the model of \citet{2005MNRAS.360..974H} 
to draw the absolute value of the kicks velocity from a one-dimensional Maxwellian 
distribution with a mean $\left<v\right>=265\,{\rm km/s}$:
\begin{equation}
    \mathcal{P}_{{\rm kick}}(v) = \sqrt{\frac{2}{\pi}}\frac{v^2}{\left<v\right>^3}\exp\left(\frac{-v^{2}}{2\left<v\right>^2}\right)
\end{equation}
and a random direction. For the uniform spherical distribution of points we employ
the algorithm of \citet{marsaglia1972}. The resulting kick vector is added to 
the Keplerian motion in the Galactic potential at the neutron stars birth position.

\subsection{The neutron star physics}
\subsubsection{The rotation evolution}
To describe the spin-down process we use the canonical lighthouse model 
\citep{1969ApJ...157.1395O}. It approximates the pulsar with magnetic dipole 
rotating in a vacuum and assumes that the total loss of the rotation energy is 
emitted in the electromagnetic spectrum. This leads to the following relation 
between magnetic field induction, period and period derivative (for details refer 
to \citet{1986bhwd.book.....S} chapter 10.5):
\begin{equation}
B = \left({\frac{3Ic^3 P \dot{P}}{8\pi^2 R^6_{{\rm NS}}}}\right)^{1/2}= \eta (P\dot{P})^{1/2}
\label{eq:BPPdotRealtion}
\end{equation} 
for a perpendicular rotator, where $\eta\simeq 3.2 \times 10^{19}\,{\rm G}\,{\rm s}^{-1/2}$.

\subsubsection{The initial parameters of pulsars}
We adopted initial spin period distribution and magnetic field strength from the 
optimal model by \citet{2006ApJ...643..332F}. In case of the period it is a 
positive normal distribution (we redraw negative values) centred at 
$\widehat{P}_{\rm init}$ and with standard deviation $\sigma_{P_{\rm init}}$. 
We initialise the magnetic field strength with values drawn from log-normal 
distribution centred at a value of $\log(\widehat{B}_{\rm init})$ and with 
standard deviation $\log(\sigma_{B_{\rm init}})$.
All four variables presented above are used to parametrize the evolution model.
We list their limits in the Table \ref{tab:ParametersConstraints}.
\subsection{Radio Properties}

\subsubsection{The minimal detectable flux}
We follow identical prescription as \citet{2011MNRAS.413..461O} to model radio selection effects.
The minimal detectable flux of a pulsar is described by the radiometer equation
\citep{1985ApJ...294L..25D} adjusted for pulsating sources: 
\begin{equation}
S_{{\rm min}}=\frac{\iota \left({\rm S/N}\right)_{{\rm min}}T_{{\rm sys}}  }{G\sqrt{n_{{\rm p}} t_{{\rm i}} \Delta f  }}\sqrt{\frac{W_{\rm e}}{P-W_{\rm e}}}
\end{equation} 
where the $\iota$ is a value describing {\em system loss}, $T_{\rm sys}$ is
the system temperature, $G$ is the gain, $n_{\rm p}$ represents the number
of polarizations, ${\Delta}f$ is the bandwidth, $t_{\rm i}$ is the integration 
time, $({\rm S/N})_{\rm min}$ represents the minimal signal to noise ratio, 
$W_{\rm e}$ is the effective width of the pulse and $P$ is the pulsar spin period. 
We supply the formula with values appropriate to the Parkes Multibeam Survey 
(see Table~\ref{tab:ParkesMBParams}).
For the system temperature $T_{\rm sys}$ we consider only the sky temperature 
$T_{\rm sky}$ in the direction of the measurement and the receiver noise 
temperatures $T_{\rm rec}$:
\begin{equation}
T_{{\rm sys}}=T_{{\rm rec}} + T_{{\rm sky}}
\end{equation}
The effective width of the pulse $W_{\rm e}$ is a function of the intrinsic 
width $W_{\rm i}$, the sampling time $\tau_{\rm samp}$, the pulsar 
dispersion measure ${\rm DM}$, the diagonal dispersion measure ${\rm DDM}$ 
(characteristic to the survey) and the interstellar scattering time
$\tau_{{\rm scatt}}$ describing the pulse widening due to the multipath
propagation (dissipation of the signal by the free electron clouds in the Galaxy).
The effective width $W_{\rm e}$ formula takes form of:
\begin{equation}
W_{{\rm e}}^2 = W_{{\rm i}}^2 +\tau_{{\rm samp}}^2 +\left(\tau_{{\rm samp}} \frac{{\rm DM}}{{\rm DDM}}\right)^2 +\tau_{{\rm scatt}}^2
\end{equation}
We obtained  the interstellar scattering time $\tau_{\rm scatt}$ using the 
model developed by \citet{2004ApJ...605..759B} in which $\tau_{{\rm scatt}}$ 
is a function of the dispersion measure ${\rm DM}$.
The minimal flux $S_{{\rm min}}$, the effective width $W_{\rm e}$, the 
system  temperature $T_{{\rm sys}}$ and $\tau_{{\rm scattering}}$ were
calculated using the  functions from the 
\text{PSREVOLVE}\footnote{\url{http://astronomy.swin.edu.au/~fdonea/psrevolve.html}} 
code developed at the Centre for Astrophysics and Supercomputing, Swinburne 
University of Technology.

%%%%%%%%%%%%%%%%%%%%%%%%%%%%%%%%%%%%%%%%%%%%%%%%%%

% Don't change these lines
\bsp	% typesetting comment
\label{lastpage}
\end{document}